\documentclass[onecolumn]{aastex6}

\newcounter{magicrownumbers} 
\newcommand\rownumber{\stepcounter{magicrownumbers}\arabic{magicrownumbers}}

\slugcomment{Submitted to The Astrophysical Journal}
\shorttitle{The Secular Mass Accretion Rate of T Pyxidis} 
\shortauthors{Godon et al.}
\watermark{ApJ Accepted}
\setwatermarkfontsize{1.0in}

\begin{document}

\title{{\bf 
The Long-Term Secular Mass Accretion Rate of the Recurrent Nova T Pyxidis
\footnote{
This research was based on observations made with the NASA/ESA Hubble 
Space Telescope, obtained at the Space Telescope Science Institute, located
in Baltimore, Maryland, USA, 
which is operated by the Association of Universities for 
Research in Astronomy, Inc., under NASA contract NAS 5-26555.    
}
}}

\author{Patrick Godon\altaffilmark{1,2}, Edward M. Sion\altaffilmark{1}}
\author{Robert E. Williams\altaffilmark{3} }
\author{Sumner Starrfield\altaffilmark{4} }


\email{patrick.godon@villanova.edu}
\email{edward.sion@villanova.edu} 
\email{wms@stsci.edu} 
\email{sumner.starrfield@asu.edu} 

\altaffiltext{1}{Department of Astrophysics \& Planetary Science, 
Villanova University, Villanova, PA 19085, USA}
\altaffiltext{2}{Henry A. Rowland Department of Physics \& Astronomy,
The Johns Hopkins University, Baltimore, MD 21218, USA}
\altaffiltext{3}{Space Telescope Science Institute, 3700 San Martin Drive,
Baltimore, MD 21218, USA}
\altaffiltext{4}{School of Earth and Space Exploration, 
Arizona State University,
Tempe, AZ 85287, USA}

\begin{abstract}

We present {\it Hubble Space Telescope} ultraviolet spectroscopy of the recurrent 
nova T Pyxidis obtained more than 5 years after its 2011 outburst 
indicating that the system might not have yet reached its deep quiescent state. 
The ultraviolet data exhibit a 20\% decline in the continuum flux 
from the pre-outburst deep quiescence state 
to the post-outburst near quiescent state. We suggest that a decline across 
each recurring nova eruption might help explain the proposed 2~mag steady decline of 
the system since 1866 \citep{sch10}.   
Using an improved version of our accretion disk model as well as 
{\it International Ultraviolet Explorer} ultraviolet and 
optical data, 
and the $4.8$~kpc distance derived by \citet{sok13} (and confirmed
by \citet{deg14}), we corroborate our previous
findings that the quiescent mass accretion rate in T Pyx  is of the order
of $10^{-6}M_{\odot}$yr$^{-1}$. Such a large mass accretion rate
would imply that the mass of the white dwarf is increasing with time.  
However, with the just-release {\it Gaia} DR 2 distance of $\sim 3.3$~kpc
(after submission of the first version of this manuscript), 
we find a mass accretion more in line with the estimate of 
\citet{pat17}, of the order of $10^{-7}M_{\odot}$yr$^{-1}$. 
Our results predict powerful soft X-ray or extreme ultraviolet 
emission from the hot inner
region of the high accretion rate disk. 
Using constraining X-ray observations and assuming the accretion disk
doesn't depart too much from the standard model,  
we are left with two possible scenarios. 
The disk either emits mainly extreme ultraviolet radiation which, 
at a distance of 4.8~kpc, is completely absorbed by the interstellar
medium,  or the hot inner disk, emitting soft X-rays, is masked by 
the bulging disk seen at a higher inclination.  

\end{abstract}

\keywords{
--- novae, cataclysmic variables  
--- stars: white dwarfs  
--- stars: individual (T Pyxidis)  
}

\section{{\bf Introduction}} \label{sec:intro}



Cataclysmic variables (CVs; see \citet{war95} for a review) 
are semi-detached interacting binaries 
with periods ranging from a fraction of an hour to days. In these
systems a white dwarf (WD) primary accretes hydrogen-rich material
from a secondary star filling its Roche-lobe. If the WD has a weak
or negligible magnetic field, the matter is accreted onto the WD
by means of an accretion disk. After enough material is accreted, 
the temperature and pressure at the base of accreted layer are high 
enough to trigger the CNO burning cycle of hydrogen, and the entire layer
undergoes a thermonuclear runaway (TNR) known as the classical nova eruption
\citep{sch49,pac65,sta72}.  
CVs for which more than one nova eruption has been observed are classified
as recurrent novae (RNe).  
If the WD  
accretes more mass during quiescence than it ejects during classical
nova eruptions,
then it is believed that it might grow to reach the Chandraskhar limit
and explode as a type Ia supernova (SN Ia; \citet{whe73,nom82}). 
This is known as the single degenerate (SD) path to SN Ia, as opposed
to the double degenerate (DD) scenario in which two CO WDs in a short
period binary system merge \citep{web84,ibe84}. 
CV systems, in theory, could also lead to SN Ia via the DD channel 
if they evolve into a double WD binary in which the total mass exceeds the
Chandrasekhar limit and with a period shorter than $\sim 13$~hr
(see \citet{liv11} for a review).   
Because of their massive WDs accreting at a high rate, 
RNe in general are considered to be ideal    
SN Ia progenitor candidates (SD channel) and for that reason they   
have been studied more extensively than other CVs.

\subsection{The Prototypical Recurrent Nova T Pyxidis} 

T Pyx is a famous recurrent nova (RN), 
having erupted six times
since 1890 \citep{waa11,sch13}. 
With a recurrence time of only $\sim$20~years (though the last quiescence
interval was twice as large, see Fig.1), the theory predicts that
T Pyx must have a massive WD accreting at a high rate
\citep{sta85,yar05}. 
As it is often the case with systems that are extensively observed, 
more questions arise than are being solved, and T Pyx remains a poorly 
understood system.  

T Pyx is surrounded by a nova shell remnant of ejected matter with a radius of 
5 arcsec \citep{due79,wil82} as well as a faint outer halo twice
as large \citep{sha89}. An analysis of the expanding knots in the nova
shell \citep{sch10} suggests that fast ejected material 
from the most recent RN eruptions is catching up and colliding with 
some slow moving material previously ejected in a classical nova
eruption in 1866. 
\citet{sch10} further proposed that T Pyx was an ordinary CV until
it underwent an eruption in 1866, increasing its 
accretion rate to a value several orders of magnitude larger than expected 
for its binary period. The enhanced mass transfer is believed to be due to a 
strong wind from the secondary star, itself driven by irradiation from
the accretion-heated WD and inner disk \citep{kni00}.   
This self-sustained feedback loop between the primary and secondary
is believed to be a transient phase: as the accretion rate
has been slowly declining, T Pyx has faded $\sim2$ mag since 
its 1866 eruption \citep{sch10}.   
Some authors \citep{pat98,kni00} have suggested that  
since 1866 \citep{sch10} T Pyx might be in a
super soft-Xray source (SSS) phase.  
However, \citet{sel08} exclude the presence of quasi-steady burning on the WD
surface, and X-ray data \citep{gre02,sel08,bal10} do not provide
observational support for this scenario.  

It is clear that during the long period of quiescence between the
classical nova eruptions T Pyx is dominated by emission from its
accretion disk and that the mass accretion rate is very high
\citep{sel08,god14}.  
In order to properly model the accretion disk, it is important to consider  
the binary system parameters, as the accretion disk model is defined by the 
mass of the central star, the radius of the central star 
(giving the lower limit for the inner radius of the disk), 
the binary separation (to help assess the outer radii of the disk), 
the inclination, and of course the mass accretion rate. 
Furthermore, as the theoretical spectrum is being scaled to the
observed spectrum, the reddening as well as the distance to the 
system become the most important parameters of all.

\clearpage

\begin{figure*}
\vspace{-9.cm}
\centering  
\includegraphics[scale=0.6]{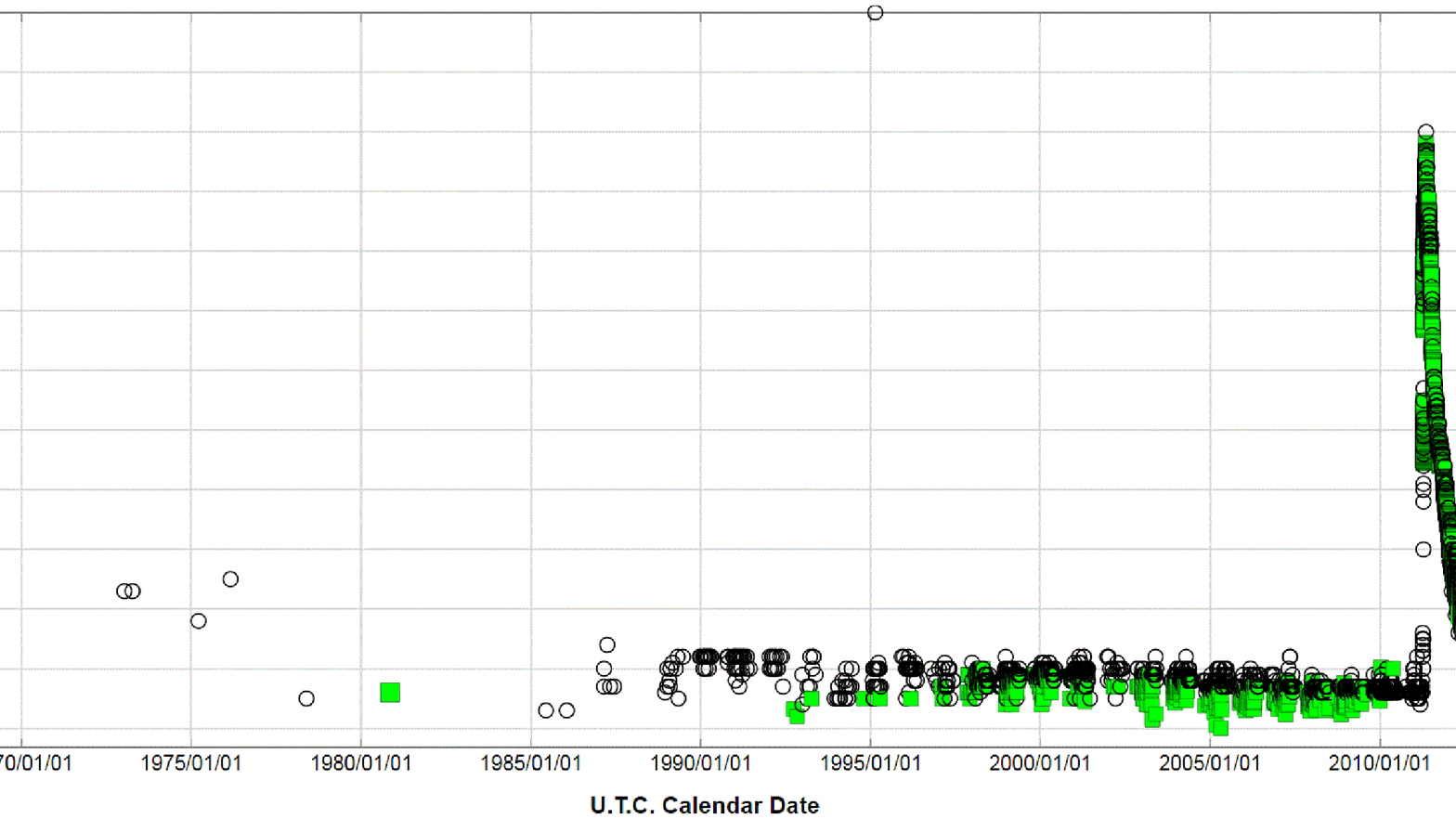}                     
\figurenum{1}
\caption{
The AAVSO light curves of T Pyx, 
from 1965 to 2018, covering the last two eruptions (in 1966
and 2011). 
The visual magnitude ($m_v$, as seen by an actual observer,
e.g. \citet{sta99,bes05}) is denoted with empty black circles, 
the standard Johnson V magnitude \citep{joh53,bes05} 
is denoted with filled green squares. 
With time, more Johnson V magnitude data have been collected   
by AAVSO members as CCDs have become easily available, 
cheaper and better.  
Both AAVSO light curves ($m_v$ and Johnson V)  
exhibit a very slight decrease in magnitude apparent from
$\sim$1988 till about 2010. 
Note that the peak magnitude is almost exactly the same ($m_v \sim 6$) 
in both eruptions. 
Note: due to maximum size, Fig.1b in the ApJ version has been
removed.  
}
\end{figure*}


\clearpage

\subsection{System Parameters}  

\begin{deluxetable*}{cccl} 
\tablewidth{0pt}
\tablecaption{T Pyxidis System Parameters} 
\tablehead{ 
Parameter    & Units       & Value        & References        
}
\startdata
$M_{\rm wd}$ & $M_{\odot}$ & 0.7-1.35     & \citet{uth10,sta85,web87,sch10}  \\
$R_{\rm wd}$ & km          & 2000-8500    & Assuming a 30,000~K WD with a mass  
$M_{\rm wd} = 1.35 M_{\odot}$ and $M_{\rm wd} = 0.7 M_{\odot}$, respectively. \\
$M_2$        & $M_{\odot}$ & 0.12-0.14    & \citet{kni00,uth10} \\  
$R_2$        & km          & 1.18$ \times 10^5$ & \citet{kni00} ($R_2=0.17 R_{\odot}$) \\  
$i$          & deg         & 10-30        & \citet{web87,uth10}. \citet{pat17} suggest $i \sim 50^{\circ}$-$60^{\circ}$ \\   
$d$          & kpc         & 4.8$\pm 0.5$ & \citet{sok13,deg14},  {\it Gaia} DR2 data implies $d\sim 3.3$~kpc.   \\  
$P$          & hrs         & 1.8295       & \citet{sch92,pat98,uth10,pat17} \\  
$E(B-V)$     &             & 0.25-0.50    & \citet{gil07,sho13,god14}  \\ 
$a$          & km          & 5-6$\times 10^5$ & The values are for $M_{\rm wd} = 0.7 M_{\odot}$, and $M_{\rm wd} = 1.35 M_{\odot}$, respectively. \\  
\enddata
\end{deluxetable*}

\paragraph{\bf{Distance}}  
Until 5 years ago the distance to T Pyx was not known and it was 
assumed to be of the order of about 1~kpc to a few kpc, 
e.g. \citet{sel95} and \citet{pat98} assumed 3.5~kpc.
More recently \citet{sok13} and \citet{deg14} 
found a distance of 4.8-5.0~kpc (respectively), and after we submitted a 
first version of this manuscript, {\it Gaia} DR2 data \citep{pru16,bro18,eye18} 
revealed a 
parallax of 0.305119199 mas with an error 0f 0.041850770 mas, 
giving a distance of 3,277~pc (between 2,882~pc and 3,798~pc).  
This is the reason why in a preliminary analysis of the archival 
{\it International Ultraviolet Explorer} ({\it IUE}) spectra of T Pyx,  
we assumed  a distance
of 1~kpc (and reddening of $E(B-V)=0.25\pm 0.05$), 
and found a mass accretion rate of only $10^{-8}M_{\odot}$yr$^{-1}$
\citep{sio10}. 
In our more recent analysis, assuming $d=4.8$~kpc (and $E(B-V)=0.35 \pm 0.05$)
we found a mass accretion rate two orders of magnitude larger \citep{god14}.  
In the present work, we initially assumed $d=4.8$~kpc, but then we took 
into account the just-released distance from {\it Gaia} DR2,  
which brings
the mass accretion of T Pyx closer to $10^{-7}M_{\odot}$yr$^{-1}$.   
This demonstrates the importance of the distance (and reddening, see below)
to determine the correct mass accretion rate of the system.  

\paragraph{\bf{Reddening}}  
At a distance of a few kpc, and with a galactic extinction of 
$E(B-V)=0.25$ 
in that direction \citep{sch98}
\footnote{\url{https://irsa.ipac.caltech.edu/applications/DUST/}}, 
one can expect the reddening towards T Pyx to be at least of that same order
of magnitude, namely $E(B-V) \sim 0.25$ or larger.  
Indeed, using co-added and merged {\it IUE} (SWP+LWP) spectra (with the
latest improved data reduction) of T Pyx and 
adopting the extinction curve of \citet{sav79}, 
\citet{gil07} obtained $E(B-V)=0.25\pm 0.02$ from the 2175~\AA\ 
dust absorption feature (``bump''). 
On the other hand, however, using diffuse interstellar bands, \citet{sho11}
found an extinction of $E(B-V) = 0.49 \pm 0.17$, basically double
that found by \citet{gil07}. Since an error of 0.05 in $E(B-V)$ (which is
quite typical for CV: \citet{ver87}) can produce
a change in the UV flux of $\sim$20\% at 3000~\AA\ and as much as $\sim$50\%
at 1300~\AA\ , the large discrepancy between the derived values of
$E(B-V)$ for T Pyxidis is frustrating. The matter is further complicated by 
the fact that deriving the reddening from the 2175~\AA\ bump  is 
accurate to within about 20\% and the extinction curve itself
is an average throughout the Galaxy and could be different in
different directions \citep{fit99}. If the reddening towards T Pyxidis
is larger than the Galactic reddening in that direction (i.e.
$E(B-V)>0.25$), then the obvious culprit is {\it obscuration} by 
the material that has
repeatedly been ejected during the recurring nova eruptions. 
Whether the `extinction properties' of the ejected material are identical
to those of the Galactic ISM is also questionable. 
It is, therefore, safe to assume for T Pyxidis a reddening value
$E(B-V)=0.35_{-0.10}^{+0.15}$.    
It is within this
context of uncertainty that in \citet{god14} we carried out 
a UV spectral analysis assuming different values for $E(B-V)$,
ranging from 0.25 to 0.50 (even though  
we derived $E(B-V)=0.35 \pm 0.05$
using the 2175~\AA\ feature from the combined and merged {\it IUE}  
and {\it Galaxy Evolution Explorer (GALEX)} spectra). 

We have updated our dereddening software and in the present work 
we carry out a new estimate of the reddening giving 
$E(B-V)=0.30 \pm 0.05$ (see Section 3.2).  
Our results for this value of $E(B-V)$ are completely consistent
with our previous results in \citet{god14}. 

\paragraph{\bf{Disk Models and Mass Accretion Rates}}  
A mass accretion rate of $\sim 10^{-8}M_{\odot}$yr$^{-1}$ 
was derived by \citet{sel08}
by scaling the flux of two \citet{wad98} disk models at  1600~\AA . 
A first model had 
$M_{\rm wd}=1.03 M_{\odot}$ with $\dot{M}=10^{-8}M_{\odot}$yr$^{-1}$, 
and the second model had  
$M_{\rm wd}=1.21 M_{\odot}$ with $\dot{M}=10^{-8.5}M_{\odot}$yr$^{-1}$. 
However, the flux at a given wavelength 
for a given mass accretion rate and a given WD mass 
cannot be scaled to much higher values of $\dot{M}$ and larger $M_{\rm wd}$
(e.g. $1.35 M_{\odot}$), 
as the 
Planckian peak moves to shorter wavelengths with increasing values
of $\dot{M}$ and $M_{\rm wd}$. 
Instead, one should rather use a realistic
accretion disk model for the values of $\dot{M}$ and $M_{\rm wd}$ in question
and fit the {\it entire wavelength range} rather than one given wavelength
(which at 1600~\AA\ doesn't include the FUV spectrum). 
Furthermore, the size of the accretion disk
in T Pyx is such that the truncation of the outer disk, either due
to tidal forcing (near $\sim 0.3a$, where $a$ is the binary separation)
or just  to the size of the Roche lobe ($\sim 0.6a$), has to be taken
into account. In our accretion disk models, the outer
disk has a lower temperature, significantly contributing to the 
UV and optical ranges, while the inner disk contributes to the EUV (especially
at the higher mass accretion rate inferred from the data). 
In the present work we take the inner and outer radius of the disk
into account to generate realistic disk models. Our models 
(see section 3, Table 4) have 
an outer disk temperature of $\sim$20,000~K and $\sim$40,000~K 
which is hotter than the $\sim$10,000~K models of \citet{wad98}
used by \citet{sel08}, and therefore have a smaller outer radius and 
provide less flux for the same $\dot{M}$ justifying our higher $\dot{M}$ results.

\paragraph{\bf{Inclination}}  
It has been shown that the inclination of the binary system is possibly low,
somewhere between 10$^{\circ}$ to 30$^{\circ}$, based on the
narrow nearly stationary emission lines \citep{web87,uth10}. 
However, \citet{pat17} suggest a higher binary inclination
$i\approx 50-60^{\circ}$, 
as inferred from {\it Hubble Space Telescope (HST)} imaging and radial velocities of the 2011
ejected shell, as well as from the soft X-ray eclipse \citep{tof13}, 
interpreting the emission lines as arising in an accretion disk wind 
\citep{sok13}.  
While, it is possible for strong X-ray modulation to occur 
even at low inclination, as is the case for  
BG CMi \citep{hel93} with $i\approx 30^{\circ}$  
(maybe due to stream disk overflow falling to smaller radii 
near the WD itself), 
we consider here also a larger inclination 
to check how this assumption affects our results
(we generate disk models assuming $i=20^{\circ}$ and $i=60^{\circ}$).  

\paragraph{\bf{The WD and Secondary}}  
Because of the short recurrence time and large $\dot{M}$ between the classical nova eruptions, 
it is expected, on theoretical grounds, that the WD in the system 
be massive, close to $1.35 M_{\odot}$ \citep{sta85,web87,sch10}. However, 
in their optical radial velocity study of emission line velocities,  
\citet{uth10} derived  a mass ratio of $q=0.20$ and an inclination of 
$i=10^{\circ}$, and, assuming a $0.14M_{\odot}$ for the secondary, 
they inferred a WD mass as low as  $0.7M_{\odot}$. 
Consequently, in the present work we consider both options: 
$M_{\rm wd} = 0.70M_{\odot}$ and $1.35M_{\odot}$.  
The radius of the WD is then taken directly from the mass-radius
relation for a non-zero temperature WD (e.g. \citet{woo95}): 
$\sim$8500~km for a $0.7 M_{\odot}$ mass, 
and $\sim$2000~km for $1.35 M_{\odot}$ mass, 
assuming that the WD has a temperature of 30,000~K. 
As to the secondary its mass is not known, but it has been estimated to be
as low as $0.06M_{\odot}$ for a high inclination \citep{pat17}, and as large
as $0.3 M_{\odot}$ (see \citet{sel08} for a review).  
In the following we will assume that the secondary mass is 
$\approx 0.13 M_{\odot}$ \citep{kni00,uth10}. We also assume that the
radius of the secondary is its Roche lobe radius. Since the secondary
mass and radius do not significantly affect the disk parameters, we do not
consider here different values.

\paragraph{\bf{The Binary Separation}}  
In order to model the disk, we  need to know its size and therefore 
the binary separation. Using Kepler's Law we find
a binary separation $a \approx 5 \times 10^5$~km assuming a 
$0.7 M_{\odot}$ WD mass, and $ a \approx 6 \times 10^5$~km 
for a $1.35 M_{\odot}$ WD mass, corresponding to $0.72 R_{\odot}$
and $0.86 R_{\odot}$ respectively.   
\\ 

In the present work, using the \citet{sok13}'s distance of 4.8 kpc  
we confirm our previously derived
mass accretion rate of $\dot{M} \sim 10^{-6}M_{\odot}$ 
\citep{god14} with our newly improved 
synthetic disk spectra to model the ultraviolet (UV) spectra of T Pyx
obtained during quiescence (pre- and post-outburst). 
However, using the {\it Gaia} DR2 data, we derive a distance of 3.3 kpc, 
which lowers the mass accretion rate to the order of 
$10^{-7}M_{\odot}$yr$^{-1}$.
We further 
extend our analysis into the optical and find that our results are
not inconsistent with the optical data.  The UV data further show that 
the mass accretion rate  has slightly decreased 4 years after the outburst 
in comparison to its pre-outburst value.  
We present the data in Section 2, give an overview of our spectral modeling
in Section 3, present the results in Section 4 and discuss them in Section 5,
and we conclude in Section 6 with a short summary.

\clearpage

\section{{\bf 
{\it IUE} Archival Data and {\it HST} Observations 
}} 

\subsection{The Pre-outburst, Outburst, and Decline UV Spectra} 

There are 58 pre-outburst {\it IUE} spectra of T Pyx which were 
collected between 1979 and 1996 \citep{szk91,gil07,sel08}
consisting of 17 spectra with   
wavelength coverage 1851-3300~\AA\  (LWP) and 
31 spectra with wavelength coverage 1150-1978~\AA\  (SWP). 
Of these, one spectrum (1979) seems to be off target, and several
others are spectra of the background sky near T Pyx. 
Consequently, in the present work we use 36 SWP spectra
together with 14 LWP spectra obtained over a period of 17 years
(1980-1996, see Table 2), exhibiting the same continuum flux level within
$\pm$9\%. The 9\% fluctuation in the continuum flux level is detected on a time
scale as small as $\sim$1 day and could be due, for example, 
to a hot spot, and/or to the stream-disk material
overflowing the disk and partially veiling the accretion disk 
at given orbital phases.
Since this fluctuation 
in flux is not very large and the
S/N of each individual spectrum is rather low, we carry out the same procedure
as in \citet{god14} and combine these spectra together
to obtain a pre-outburst spectrum of T Pyx that can be used to assess the
continuum flux level change of the post-outburst spectra. 
Details of {\it IUE} exposures are listed in Table 2 (in chronological order)
together with a {\it GALEX}
pre-outburst spectrum obtained in 2005. 

T Pyx was later observed with {\it HST}/{\it Space Telescope Imaging Spectrograph} 
(STIS) ($\sim$1150-2900~\AA ) 
first during its outburst while the spectrum was dominated by 
broad and strong emission lines with a flux level 10-100 times
larger than in quiescence  
\citep{sho13,deg14}, then with STIS and the {\it Cosmic Origins Spectrograph} (COS) 
($\sim$900-1725~\AA ) during its late decline from outburst \citep{god14,deg14}
as its flux level reached its pre-outburst value and the emission
lines almost vanished,
and more recently with COS ($\sim$900-2150~\AA )
as the system continues to decline (as reported here, see next subsection) 
with a continuum flux level
slightly below its pre-outburst spectrum.  
The observation log of the {\it HST} post-outburst exposures is listed in 
Table 3 (also in chronological order). Each {\it HST} spectrum consists 
of several exposures obtained successively. For that reason in Table 3
we mark in bold the date of the first exposure to help differentiate
between the actual spectra obtained at different epochs.    

\clearpage

\begin{deluxetable*}{ccccc}
\tablewidth{0pt}
\tablecaption{T Pyxidis Pre-Outburst UV Archival Spectra}
\tablehead{
Telescope        & Data ID   & Date (UT)  & Time (UT)&  Exp. Time     \\
                 &           & yyyy-mm-dd & hh:mm:ss &  seconds   
}
\startdata
 {\it IUE} & SWP08973            & 1980-11-05 & 04:09:56 & 12900               \\ 
 {\it IUE} & SWP29318            & 1986-09-27 & 18:22:55 & 16200             \\ 
 {\it IUE} & SWP32218            & 1987-11-02 & 12:12:44 & 16800              \\ 
 {\it IUE} & SWP32899            & 1988-02-11 & 07:27:19 & 12780              \\ 
 {\it IUE} & SWP33034            & 1988-03-03 & 04:29:13 & 23580   \\  
 {\it IUE} & SWP34696            & 1988-11-05 & 14:01:04 & 17160               \\ 
 {\it IUE} & SWP37536            & 1989-11-07 & 14:12:10 & 16800               \\ 
 {\it IUE} & SWP43442            & 1991-12-22 & 10:20:50 & 15600               \\ 
 {\it IUE} & SWP44182            & 1992-03-16 & 03:55:48 & 15000               \\ 
 {\it IUE} & SWP44948            & 1992-06-17 & 22:09:41 & 16800                \\  
 {\it IUE} & SWP46605            & 1992-12-28 & 12:15:49 & 16320   \\  
 {\it IUE} & SWP47057            & 1993-02-27 & 06:09:34 & 16500               \\ 
 {\it IUE} & SWP47323            & 1993-03-20 & 04:47:40 & 20100   \\  
 {\it IUE} & SWP47328            & 1993-03-21 & 05:30:39 & 18000   \\  
 {\it IUE} & SWP47332            & 1993-03-22 & 04:05:15 & 23100               \\ 
 {\it IUE} & SWP49365            & 1993-11-29 & 11:49:59 & 9600                \\ 
 {\it IUE} & SWP49366            & 1993-11-29 & 15:25:32 & 10680               \\ 
 {\it IUE} & SWP50099            & 1994-02-24 & 06:07:47 & 11400               \\ 
 {\it IUE} & SWP50100            & 1994-02-24 & 09:47:40 & 10800               \\ 
 {\it IUE} & SWP50596            & 1994-04-20 & 01:58:50 & 24480               \\ 
 {\it IUE} & SWP52886            & 1994-11-23 & 12:27:47 & 9000                \\ 
 {\it IUE} & SWP52887            & 1994-11-23 & 15:52:41 & 9480                \\ 
 {\it IUE} & SWP53810            & 1995-02-02 & 04:04:13 & 11400               \\ 
 {\it IUE} & SWP54590            & 1995-05-03 & 23:48:06 & 11100   \\  
 {\it IUE} & SWP54591            & 1995-05-04 & 03:30:52 & 11700   \\  
 {\it IUE} & SWP56240            & 1995-11-26 & 19:15:40 & 12000               \\ 
 {\it IUE} & SWP57030            & 1996-05-01 & 23:24:57 & 12000               \\ 
 {\it IUE} & SWP57031            & 1996-05-02 & 03:10:44 & 12600               \\ 
 {\it IUE} & SWP57032            & 1996-05-02 & 23:16:03 & 12000               \\ 
 {\it IUE} & SWP57033            & 1996-05-03 & 03:04:57 & 12000               \\ 
 {\it IUE} & SWP57034            & 1996-05-03 & 08:05:54 & 10800               \\ 
 {\it IUE} & SWP57035            & 1996-05-03 & 13:03:18 & 6600               \\ 
 {\it IUE} & SWP57039            & 1996-05-04 & 03:19:54 & 7799               \\ 
 {\it IUE} & SWP57042            & 1996-05-04 & 09:41:57 & 9899               \\ 
 {\it IUE} & SWP57047            & 1996-05-05 & 02:23:20 & 10800               \\ 
 {\it IUE} & SWP57055            & 1996-05-06 & 03:05:51 & 10800               \\ 
 {\it IUE} & LWR07724            & 1980-11-05 & 02:08:20 &  7200               \\ 
 {\it IUE} & LWP09204            & 1986-09-27 & 16:15:42 & 7200                \\ 
 {\it IUE} & LWP11996            & 1987-11-02 & 17:00:13 & 6600                \\ 
 {\it IUE} & LWP12644            & 1988-02-11 & 05:20:34 & 7200                \\ 
 {\it IUE} & LWP12791            & 1988-03-03 & 07:29:37 & 12780   \\  
 {\it IUE} & LWP14383            & 1988-11-05 & 11:44:47 & 7800                \\ 
 {\it IUE} & LWP16757            & 1989-11-07 & 11:52:47 & 7800                \\ 
 {\it IUE} & LWP22052            & 1991-12-22 & 14:45:26 & 7200                \\ 
 {\it IUE} & LWP22608            & 1992-03-15 & 08:09:57 & 9600                \\ 
 {\it IUE} & LWP23317            & 1992-06-18 & 02:54:33 & 6900                \\ 
 {\it IUE} & LWP24612            & 1992-12-28 & 09:48:28 & 8400    \\  
 {\it IUE} & LWP25020            & 1993-02-27 & 10:55:54 & 6600                \\ 
 {\it IUE} & LWP32286            & 1996-05-05 & 13:37:15 & 5700                \\ 
 {\it IUE} & LWP32287            & 1996-05-06 & 11:26:01 & 6600                \\ 
{\it GALEX} & GI2\_023004\_T\_PYX  & 2005-12-20 & 03:30:05 & 880             \\ 
\enddata
\end{deluxetable*}

\begin{deluxetable*}{ccccccc}
\tablewidth{0pt}
\tablecaption{T Pyxidis Post-Outburst {\it HST} Spectra}
\tablehead{
{\it HST}  & Filters  & Central & Data ID & Date (UT) & Time (UT)& Exp.Time     \\
Instrument & Gratings & Wavelength &     & yyyy-mm-dd & hh:mm:ss &  seconds   
}
\startdata
 STIS      & E140M    & 1425  & obg101010 & {\bf{2011-05-07}}& 08:01:44 &  571   \\
 STIS      & E230M    & 1978  & obg101020 &      2011-05-07  & 08:26:53 &  571   \\
 STIS      & E140M    & 1425  & obg199010 & {\bf{2011-07-28}}& 18:52:15 &  285   \\
 STIS      & E230M    & 1978  & obg199020 &      2011-07-28  & 19:03:07 &  285   \\
 STIS      & E140M    & 1425  & obg103010 & {\bf{2011-10-04}}& 12:56:11 &  285   \\
 STIS      & E230M    & 1978  & obg103020 &      2011-10-04  & 13:14:18 &  285   \\
 STIS      & E140M    & 1425  & obx701010 & {\bf{2012-03-28}}& 15:25:23 & 2457   \\
 STIS      & E140M    & 1425  & obx701020 &      2012-03-28  & 16:50:44 & 3023   \\
 COS       & G130M    & 1055  & lbx702ouq &      2012-03-28  & 18:36:38 &  461   \\
 COS       & G130M    & 1055  & lbx702owq &      2012-03-28  & 18:47:22 &  461   \\
 COS       & G130M    & 1055  & lbx702oyq &      2012-03-28  & 18:58:06 &  461   \\
 COS       & G130M    & 1055  & lbx702p0q &      2012-03-28  & 19:08:50 &  461   \\
 STIS      & E140M    & 1425  & obxs01010 & {\bf{2012-12-21}}& 05:18:19 & 2449   \\
 STIS      & E140M    & 1425  & obxs01020 &      2012-12-21  & 06:43:06 & 3015   \\
 COS       & G130M    & 1055  & lbxs02010 &      2012-12-21  & 08:31:37 & 1744   \\
 STIS      & E140M    & 1425  & obxs03010 & {\bf{2013-07-26}}& 08:36:35 & 2449   \\
 STIS      & E140M    & 1425  & obxs03020 &      2013-07-26  & 09:57:04 & 3015   \\
 COS       & G130M    & 1055  & lbxs04010 &      2013-07-26  & 11:46:49 & 1744   \\
 COS       & G140L    & 1105  & lcue01010 & {\bf{2015-10-13}}& 00:00:16 &  788   \\
 COS       & G130M    & 1055  & lcue01020 &      2015-10-13  & 01:20:11 & 3466   \\
 COS       & G140L    & 1105  & lcue02010 & {\bf{2016-06-01}}& 02:14:18 &  788   \\
 COS       & G130M    & 1055  & lcue02020 &      2016-06-01  & 03:41:58 & 3466   \\
\enddata
\end{deluxetable*}

\clearpage

The first 6 {\it HST} spectra (obtained between May 2011 and July 2013) 
were presented and analyzed in \citet{god14} and \citet{deg14} and showed that
by December 2012 the continuum flux level had reached its 
pre-outburst value (see Fig.2), and remained constant thereafter 
(the July 2013 spectrum was basically identical to the December 2012
spectrum). The AAVSO light curve of the system also indicated 
that by $\sim$2013 the light curve had reached its quiescence level.
The system, therefore, seemed to have returned to 
exactly the same quiescent state. The Dec 2012 and Jul 2013  
post-outburst {\it HST} spectra
agree very well with the pre-outburst {\it IUE} and {\it GALEX} spectra, as shown 
in Fig.3 (except for detector edge noise in {\it GALEX} and {\it IUE}). 
As a result of this, in \citet{god14} we modeled the combined {\it HST} and {\it IUE} 
spectra with an accretion disk, yielding a  mass accretion rate 
$\dot{M} \approx 2 \pm 1 \times 10^{-6} M_{\odot}$yr$^{-1}$.

\begin{figure*}
\figurenum{2}
\vspace{-4.cm} 
\plotone{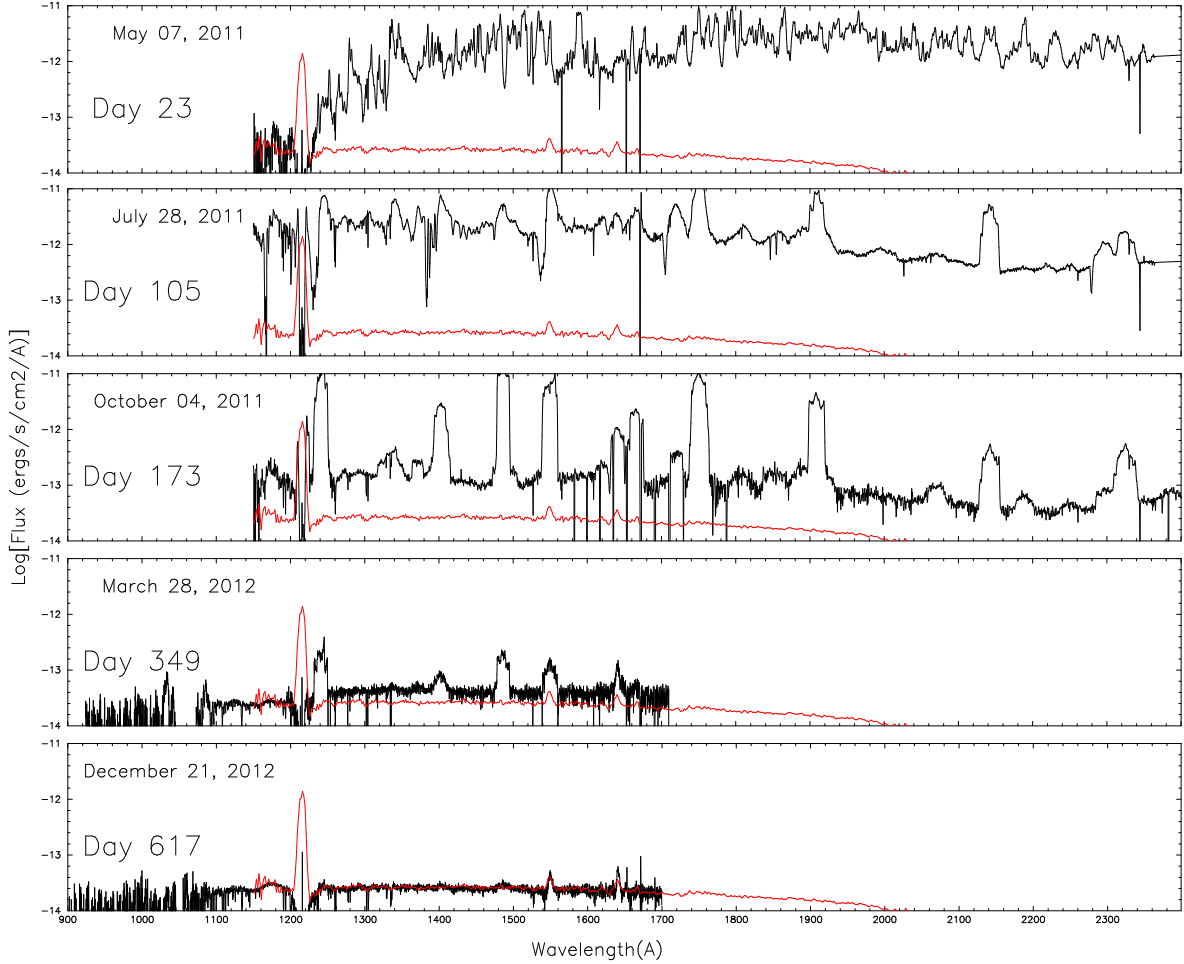}
\vspace{1.cm} 
\caption{T Pyxidis decline from outburst.  
The first five {\it HST} spectra of T Pyx (as listed in Table 3)  
obtained as the system declined from its original eruption 
are shown in chronological order from top to bottom (solid black lines).
The flux is given in log of erg~s$^{-1}$cm$^{-2}$\AA$^{-1}$
and the wavelength is in \AA .  
The observation date and the 
number of days since eruption are indicated on the left in each panel.    
The pre-eruption {\it IUE}  spectrum (generated from the co-added and  
combined {\it IUE}  datasets listed in Table 2) is shown in red for 
comparison. The {\it HST} December 21, 2012 observations reveal a UV 
spectrum almost identical to the {\it IUE}  pre-eruption spectrum (bottom panel). 
}
\end{figure*}

\begin{figure}
\figurenum{3}
\vspace{-3.cm} 
\plotone{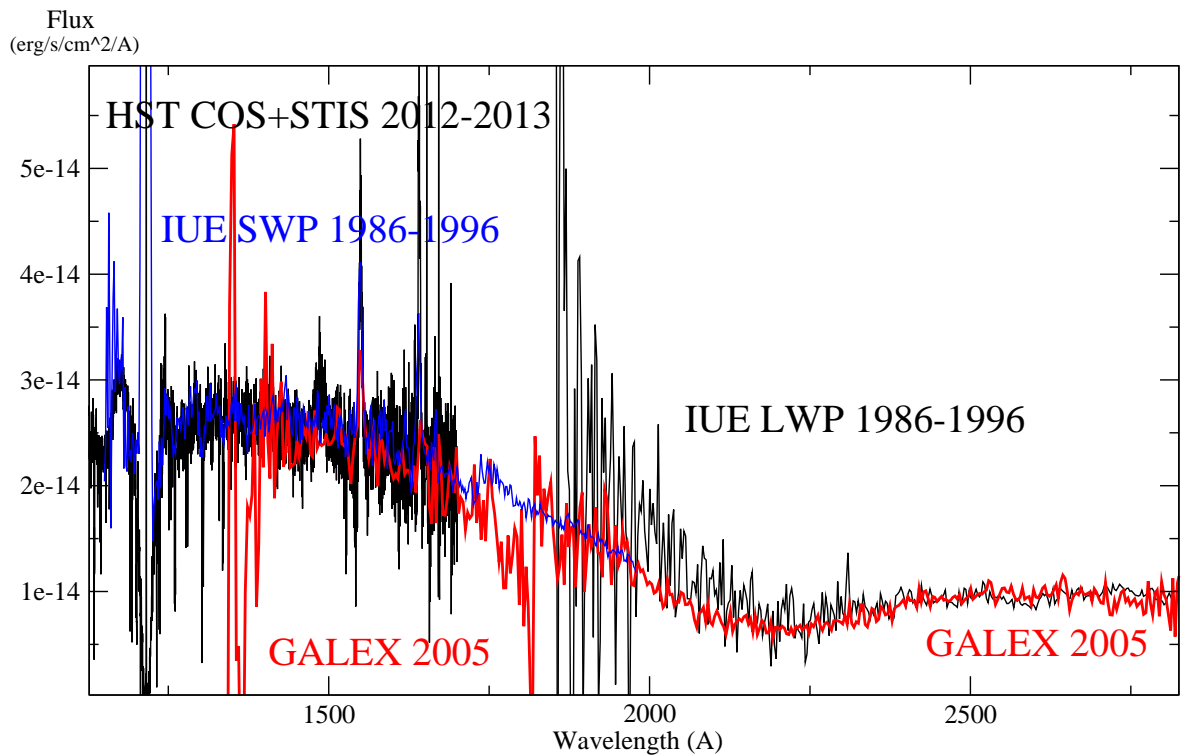}
\vspace{-1.cm} 
\caption{
The pre-eruption combined (SWP and LWP) {\it IUE}  spectra  are shown together
with the {\it GALEX} spectrum and the post-eruption {\it HST} spectra.
All the spectra are color-coded and the observations years are indicated  
as marked. 
For a better contrast, and since they do not overlap, 
both the HST spectrum 
(in the short wavelengths) and IUE LWP spectrum (in the longer wavelengths)
have been set to black. 
The S/N of the {\it IUE}  LWP spectrum is very low and deteriorates toward
the edges (as seen below 2000~\AA ). 
The {\it GALEX} spectrum (two spectral segments in red) presents 
a similar low S/N at its edges (including where the two segments 
meet around 1700-2000~\AA ). 
In spite of this, all the spectra agree  relatively well in the
regions where the S/N is good. 
}
\end{figure}

\clearpage

\subsection{The 2015 \& 2016 Late Decline/Quiescent {\it HST}/COS Spectra}

We obtained two additional {\it HST}/COS 
far ultraviolet (FUV) spectra: the first one more than two years after the last {\it HST} spectrum, 
namely in October 2015, and the second one in June 2016. 

For each two-orbit observation,  
we used two different FUV COS configurations - each taking one orbit
and in time tag mode. 
In the first orbit, COS was configured 
with the G140L grating centered at 1105~\AA, from which
we extracted one spectral segment from 1150~\AA\ to 2150~\AA. 
In the second orbit, we used the G130M grating 
centered at 1055~\AA , 
from which we extracted two spectral segments 
from 900~\AA\ to 1045~\AA\ and from 1055~\AA\ to 1200~\AA .  
The data were collected
through the primary science aperture (PSA) of 2.5 arcsec diameter, 
and processed with CALCOS version 3.1.8 \citep{hod07,hod11} 
through the pipeline, 
which produces, for each {\it HST} orbit, four sub-exposures generated
by shifting the position of the spectrum on the detector by 
20~\AA\ each time. This strategy is used to reduce detector effects
associated with COS. A description of the COS four  
positions sub-exposures and detector artifacts can be found 
in \citet{god17a}.  
For the first orbit (COS G140L 1105) the effective good exposure time 
of each sub-exposure was a little less than $\sim$200 sec.
For the second orbit (COS G130M 1055) the exposure time
for each sub-exposure was $\sim 866$ sec.    
The resulting spectrum covers the Lyman 
series down to its cut-off wavelength, including the L$\alpha$ region.   
Due to detector
artifacts (especially near the edges), the resulting spectrum 
is  not very reliable 
in the region where the segments overlap (1150-1200~\AA ) and in the shortest
wavelengths ($< 1000$~\AA , see Figs.4a \& b). 

The October 2015 spectrum exhibits a net drop (of $\sim 20$\%) 
in the continuum flux level
when compared to the Dec 2012 - Jul 2013 {\it HST} spectra 
and {\it IUE}  pre-outburst spectrum (see Fig.4a). The June 2016 spectrum 
has a continuum flux slightly higher than
the October 2015 spectrum, but still well below the pre-outburst level 
(see Fig.4b).  

To better emphasize the continuum flux changes, we average the spectra
over the wavelength region 1400~\AA\ - 1700~\AA, omitting the emission
lines. Namely, the averaged flux is given by   
$$ 
\frac{1}{\lambda_2-\lambda_1}
\int_{\lambda_1}^{\lambda_2} F_{\lambda} d\lambda 
$$  
with 
$\lambda_1=1400$~\AA\ and $\lambda_2=1700$~\AA .
We present the averaged UV flux of the pre-outburst ({\it IUE}  +{\it GALEX}) and post-outburst
({\it HST}) spectra in Fig.5. 
The first obvious behavior is 
the 9\% fluctuation 
in the flux level seen in the {\it IUE} data and 
present even on a time-scale of $\sim 1~$day
(year 1996). The {\it GALEX} flux data point, obtained at the end of 2005, is 
rather low, but not inconsistent with the {\it IUE} data. The {\it HST} 
post-eruption data show that the Dec 2012 and Jul 2013 data points fall 
within the range of values of the pre-outburst {\it IUE} data and explain 
how they were interpreted as the system having come back to its exact
pre-eruption quiescent state \citep{god14}. Only the October 2015 and June 2016 
data points reveal that the system has apparently not yet reached quiescence 
and the UV continuum flux level is still decreasing. The 2016 data point
seems to bounce back compared to the 2015 data point, however it can be
understood as a `normal' modulation of the continuum flux level, namely
within the 
$\pm 9$\% 
fluctuation. 
Both the 2015 and 2016 data points 
(with $F_{\lambda} \approx 2 \times 10^{-14}$erg~s$^{-1}$cm$^{-2}$\AA$^{-1}$) 
are definitely below the averaged flux level
of the pre-outburst data   
(with 
$F_{\lambda} \approx 2.5 \times 10^{-14}$erg~s$^{-1}$cm$^{-2}$\AA$^{-1}$),
showing a drop of $\sim 20$\%.
At this stage, it is not clear whether the system has reached quiescence
or will decline further.

\clearpage

\begin{figure}
\vspace{-5.cm} 
\figurenum{4a}
\plotone{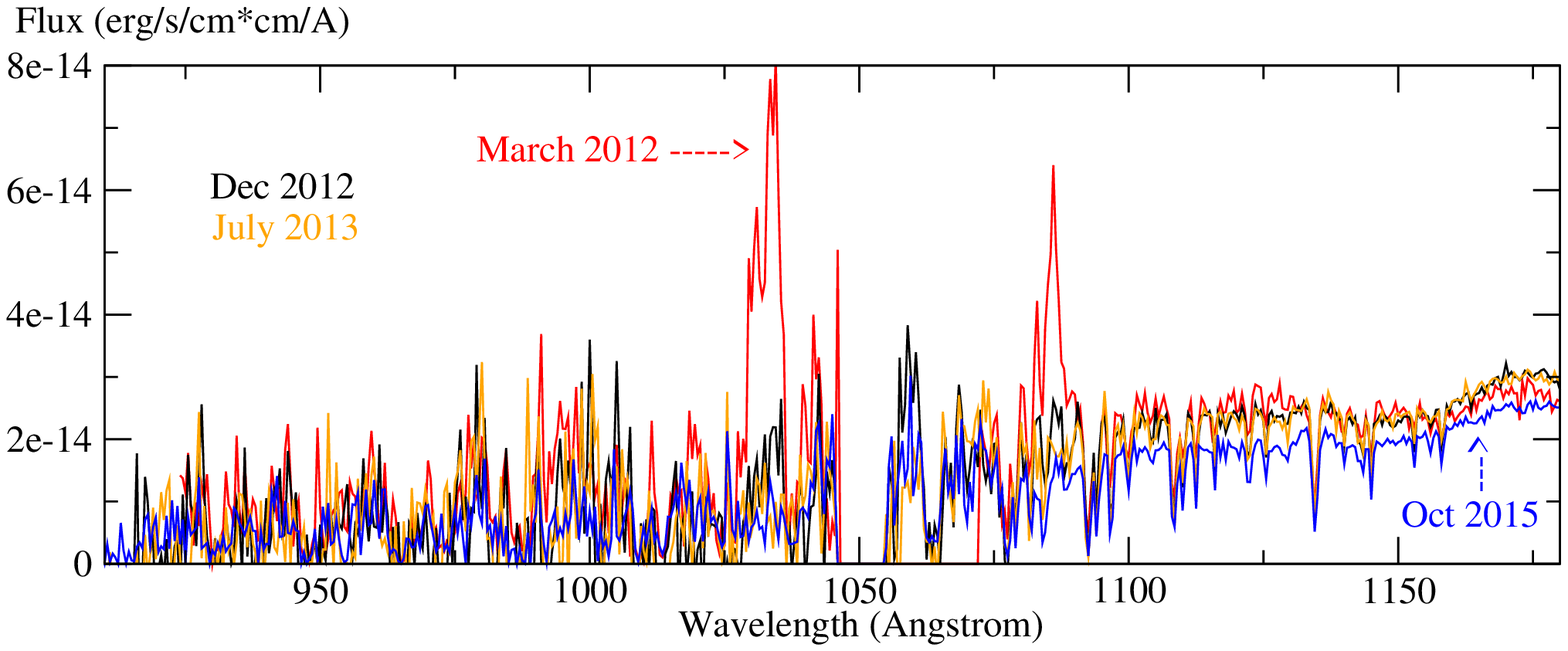}
\vspace{-0.5cm} 
\plotone{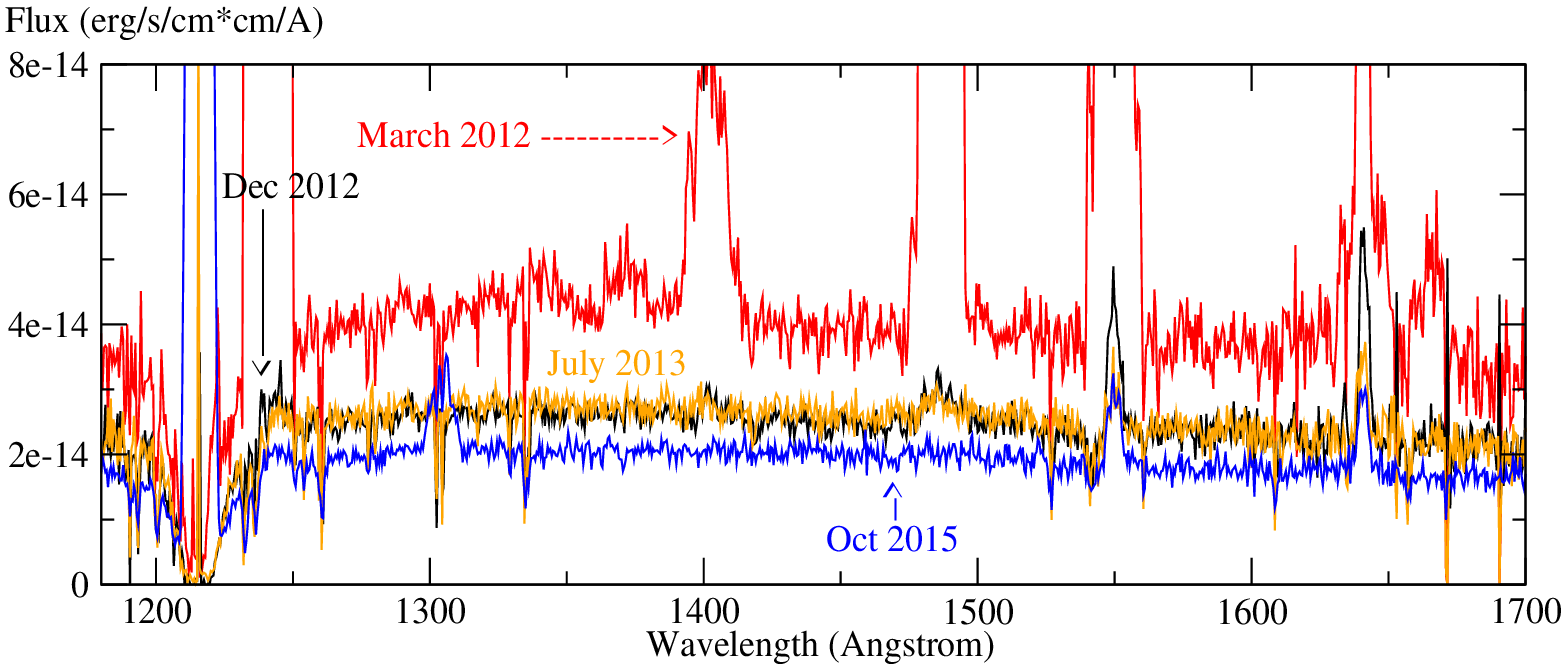}
\vspace{-18.cm} 
\caption{
The 2012, 2013 and 2015 {\it HST} FUV spectra of T Pyxidis 
are displayed showing the 
late decline of the system into quiescence.
The first panel shows the COS G130M (1055~\AA)  
spectra going down to the Lyman limit, but with a very low S/N 
(some of the flux below 1000~\AA\ 
is actually negative, a sign that it is unreliable). 
In the short wavelengths the spectra have about the same continuum flux
level, except for the March 2012 spectrum exhibiting a couple 
strong emission lines. The Oct 2015 spectrum has a slightly lower
continuum flux level above 1000~\AA . 
In panel (b) we  show three STIS E140M (centered at 1425~\AA) and one 
COS G140L (centered at 1105~\AA) spectra (see Table 3).
The March 2012 spectrum shows strong emission lines and has a 
continuum flux level significantly larger than the other spectra. 
The Oct 2015 spectrum shows a net decline in the continuum flux level. 
For most part, the Jul 2013 and Dec 2012 spectra are indistinguishable.  
The epochs are color-coded and indicate a further decrease in 
flux below the pre-eruption level for the data obtained
after 2013.  
Note that the fluxes in the region where the STIS and COS spectra
overlap ($\sim 1150$-1200~\AA\ ) do not agree,  
especially for the March 2012 dataset.  
}
\end{figure}

\clearpage

\begin{figure}
\vspace{-5.cm} 
\figurenum{4b}
\plotone{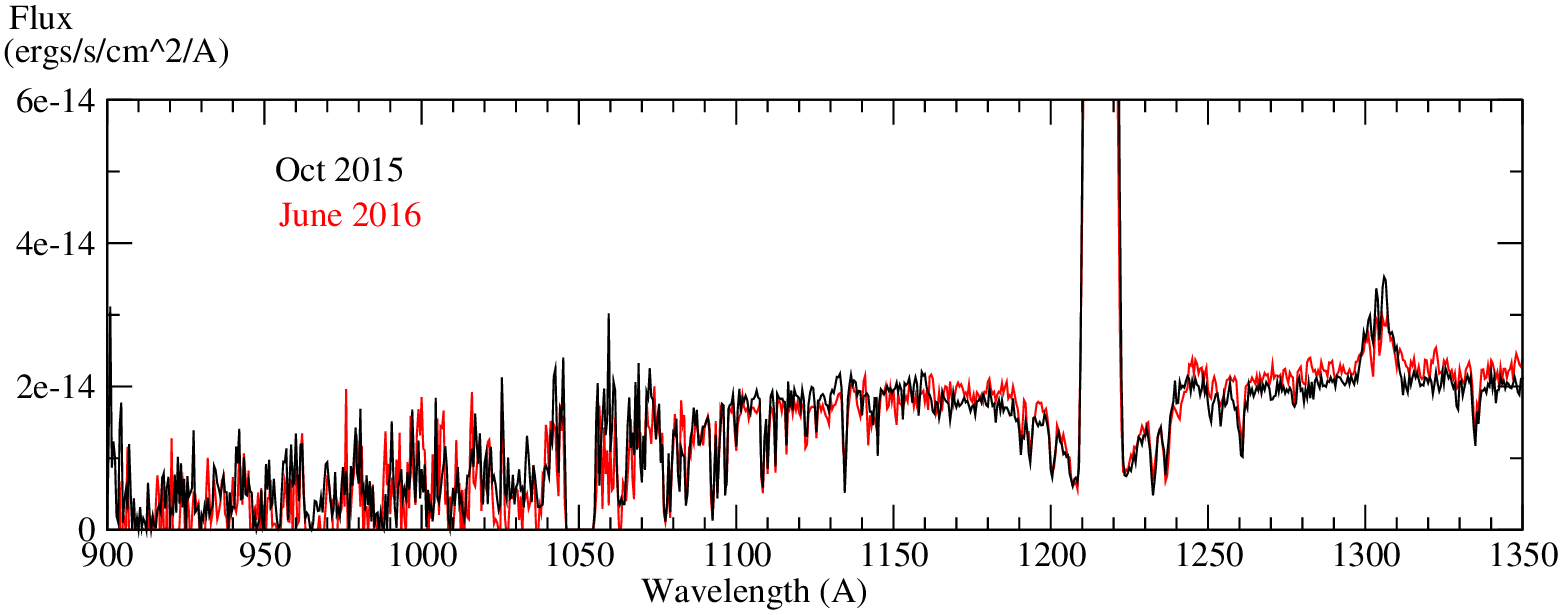}
\vspace{-0.5cm} 
\plotone{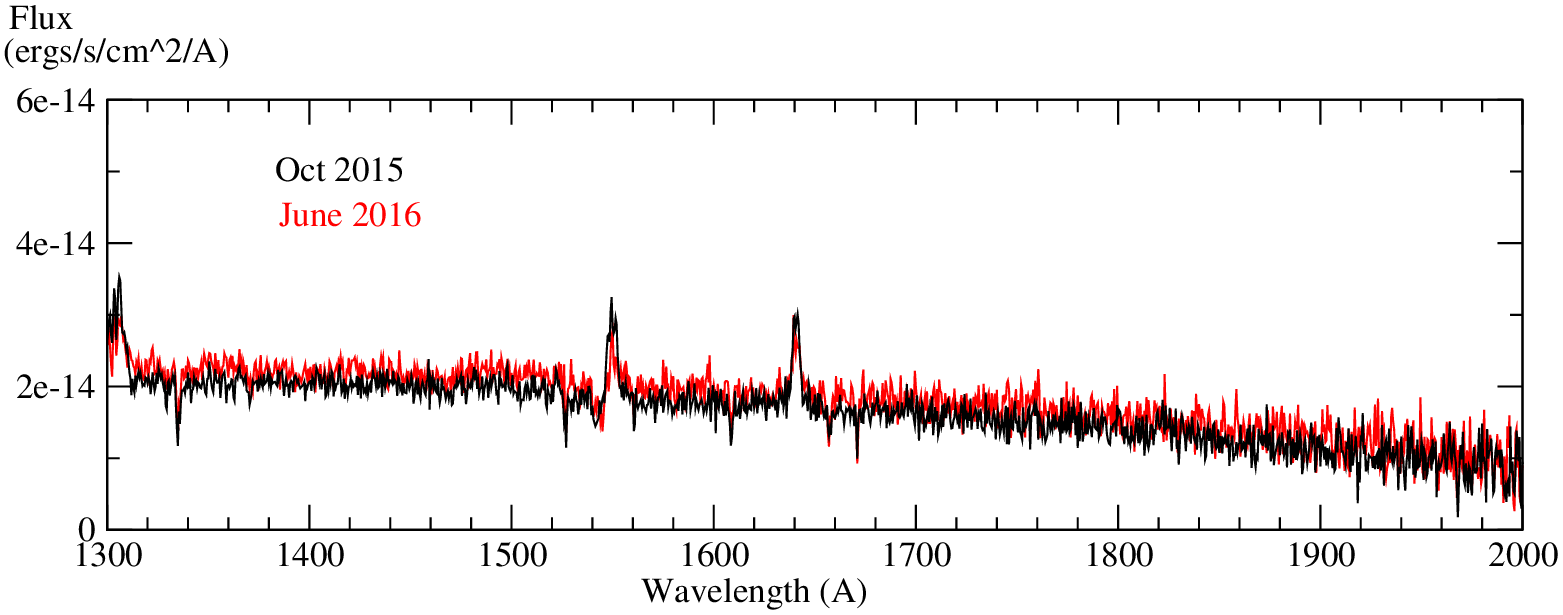}
\vspace{-18.cm} 
\caption{
The June 2016 HST/COS spectrum of T Pyxidis is shown (in red) 
together with the October 2015 spectrum (in black) 
and displays a slight increase in flux especially in the longer
wavelengths.  
}
\end{figure}

\clearpage

\begin{figure}
\figurenum{5}
\vspace{-3.cm} 
\plotone{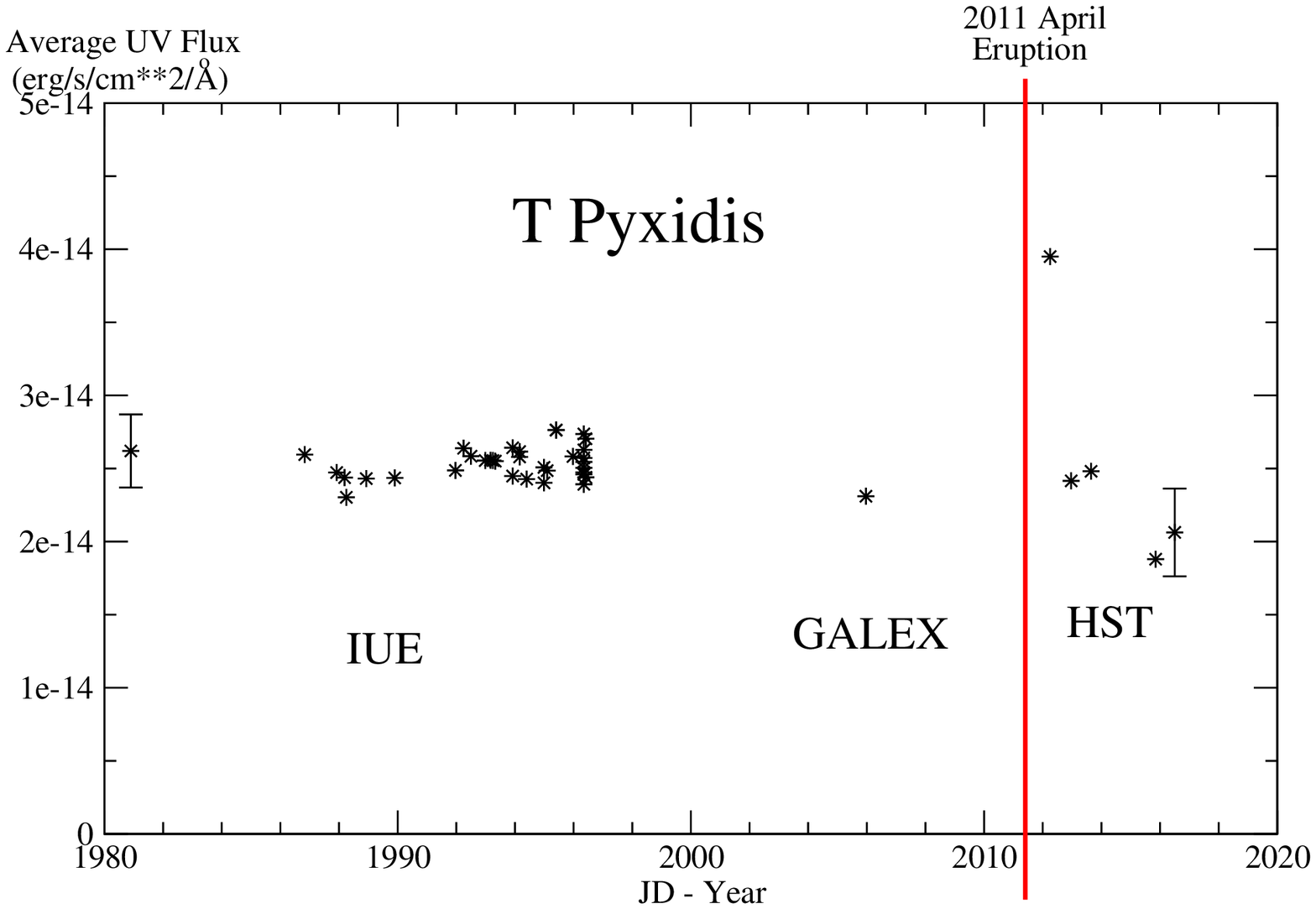}      
\vspace{-1.cm} 
\caption{
The average flux for the {\it IUE} SWP, {\it GALEX} and {\it HST}/STIS spectrum are shown
as a function of time (see Tables 2 and 3 for exact dates). The 
flux was averaged 
over the spectral region $\sim 1400$ to $\sim 1700$~\AA\ , 
excluding emission lines. 
For clarity the error bar (in $F_{\lambda}$) has been drawn only 
for the first and last data points and represents the average value
of the flux error around 1400~\AA\ (20 to 25\%). 
{\it IUE} data points obtained over a period of 
a few days in 1996 reveal that the flux can vary by as much as 
$\sim 18$\% 
($\pm $9\% 
from the median value).  The 2005 
{\it GALEX} and 2012-2013 {\it HST} data points appear within the  flux fluctuation of
the {\it IUE} pre-eruption data points. Only the 2015 and 2016 data, obtained
more than $4\frac{1}{2}$ years after the eruption (vertical red line), show a
drop below the pre-eruption level. 
}
\end{figure}

\clearpage

\section{{\bf
Modeling 
}} 

Our spectral modeling tools and technique have been previously 
described extensively  in numerous works such as \citet{god12,god16}, 
as a consequence, we limit ourselves here to a brief description of the 
modeling but include a comprehensive account of the recent improvements
we made over our previous UV spectral analysis of T Pyx
in \citep{god14}. 

\subsection{Modeling of the Accretion Disk Spectrum} 
For the disk we assume the standard model 
\citep{sha73,pri81}, namely,  
the disk is optically thick, it has a negligible vertical thickness $H/r << 1$,
it is axisymmetric and the energy dissipated between shearing
adjacent annuli of matter is radiated locally in the $\pm$z directions. 
As a consequence, the temperature is solely a function of the radius $r$, 
and is completely defined by the mass $M_{\rm wd}$ of the central star, the
mass accretion rate $\dot{M}$, and the inner radius of the disk $R_{\rm in}$. 
The luminosity and spectrum of the disk are obtained by integrating
over the entire surface area of the disk from the inner radius $R_{\rm in}$
to the outer radius $R_{\rm out}$.   

In order to generate a disk spectrum, 
we divide the disk into $N$ rings with radius $r_i (i,1,2..N)$
each with a temperature $T(r_i)$ given by the standard disk model and 
defined when $M_{\rm wd}$, $\dot{M}$, and $R_{\rm in}$ are given.  
For each ring, a synthetic spectrum is generated by running 
Hubeny's synthetic stellar atmosphere suite
of codes TLUSTY, SYNSPEC and ROTIN \citep{hub88,hub95}. A final
disk spectrum is then obtained by running DISKSYN, which 
combines the rings spectra together for a given
inclination and takes into account Keplerian broadening and limb darkening
\citep{wad98}. TLUSTY \& SYNSPEC are described in detail in 
\citet{hub17a,hub17b,hub17c}.  

Our present modeling includes 
a number of improvements we recently made as follows.  

\paragraph{\bf{The Inner Disk Radius}}  
The standard disk model, \citet{wad98}'s disk models, and our
previous disk models all assume that the inner radius of the
disk corresponds to the radius of the central star
$R_{\rm in}=R_{\rm wd}$.
In other words, the radial thickness of the 
boundary layer is negligible: $\delta_{\rm BL} << R_{\rm wd}$
(where the size of the boundary layer is determined by the 
vanishing of the shear $d \Omega /dR=0$ at $R_{\rm in} =R_{\rm wd} + \delta_{\rm BL}$
\citep{pri81}, in cylindrical coordinates $(R,\phi,z)$). 
This assumption is valid for 
an optically thick boundary layer when the mass accretion rate
is moderately large (here $\dot{M} \approx 10^{-9}-10^{-7}M_{\odot}$),
and negligible in comparison to the Eddington limit 
($\dot{M} << \dot{M}_{\rm Edd}$). 
For a very large mass accretion rate, the size of the 
optically thick boundary layer increases \citep{god96,god97}
and one has $\delta_{\rm BL} \approx 0.1-0.5 R_{\rm wd}$ or so. 
When the boundary layer is optically thin (which usually happens
at low mass accretion rates), its size also increases
\citep{pop93,pop99}.     

In the present work we take the inner radius of the accretion
disk to be larger than the radius of the star ($R_{\rm in}> R_{\rm wd}$) 
to accommodate for 
the possible presence of an extended boundary layer.  
This also allows us to consider a heated WD with an inflated radius 
in comparison to the zero-temperature radius, or a disk that is
truncated by the WD magnetic field. The important point here is
that we do not generate a disk model with an inner radius 
at $R_{\rm wd}$ and truncate it at $R_{\rm in}$, but, rather, we generate
a disk model with the inner radius at $R_{\rm in} > R_{\rm wd}$. 
Since the no-shear ($d \Omega/dR=0$) boundary condition is 
imposed at the disk inner radius (rather than the WD radius) 
this difference produces  
a disk that is colder in its inner region relatively to the standard disk model
(it is a standard disk model with $R_{\rm wd}$ replaced by 
$R_{\rm in} > R_{\rm wd}$). 
Namely, the disk radial temperature profile \citep{pri81}
can be written as 
\begin{equation}
T_{\rm eff}(x)= T_0 x^{-3/4} (1-x^{-1/2})^{1/4}, 
\end{equation}
with 
\begin{equation}
T_0=64,800~{\rm K} \times \left[  
\left( \frac{ M_{\rm wd}}{M_{\odot}} \right)
\left( \frac{ \dot{M}}{10^{-9}M_{\odot}{\rm yr}^{-1}} \right)
\left( \frac{ R_{\rm in} }{ 10^9{\rm cm}} \right)^{-3}
\right]^{1/4},  
\end{equation}  
and where $x=R/R_{\rm in}$ \citep{wad98}, but now $R_{\rm in}$ 
is the inner radius of the disk \citep{god17b}.   
The maximum disk temperature, $T_{\rm max} = 0.488 T_0$ reached at 
$x=1.36$, is given by 
\begin{equation}
T_{\rm max} =177,826~{\rm K} 
\left( \frac{ M_{\rm wd}}{M_{\odot}} \right)^{1/4} 
\left( \frac{ \dot{M}}{10^{-6}M_{\odot}{\rm yr}^{-1}} \right)^{1/4} 
\left( \frac { R_{\rm in} } { 10,000~{\rm km}}
\right)^{-3/4},  
\end{equation}
where for convenience we have now written $\dot{M}$ in units of $10^{-6}M_{\odot}$yr$^{-1}$,
which is the order of magnitude of the mass accretion we obtained for T Pyx in \citet{god14}.   
It is then apparent that for 
a $\sim 0.7 M_{\odot}$ WD with a radius of $\sim 8,500$~km (see Section 1.2), 
one has $T_{\rm max} \approx 150,000$~K, while for  
a $\sim 1.35 M_{\odot}$ WD with a radius of $\sim 2,000$~km, $T_{\rm max} \approx 500,000$~K. 

We note, however, that the maximum disk temperature achieved in the inner 
region can be decreased by increasing $R_{\rm in}$. 
Our  modified accretion disk models were already 
presented and used in \citet{god17b} and \citet{dar17}, and further details
can be found therein. 

\paragraph{\bf{The Outer Disk Radius}}  

In our previous work \citep{god14}, based on the work of \citet{wad98},
the disk was extended   to a radius where the temperature reached 10,000~K.
Such a radius, 
depending on the binary separation, mass of the WD, 
mass accretion rate, and state of the system,
could be larger or smaller than the actual radius of the disk.  
We improved our modeling by choosing the outer disk radius 
to represent the physical radius of 
the disk, which can now be extended to include outer rings with a
temperature as low as 3,500~K.  

Due to the tidal interaction of the secondary star, 
the size of the disk is expected to be between $0.3a$ (where $a$ is
the binary separation) for a mass ratio $q=M_2/M_1 \approx 1$, 
and about $0.6a$ for $q << 0.1$ 
\citep{pac77,goo93}. T Pyx, with a binary mass
ratio of $q = 0.20 \pm 0.03$  
(\citet{uth10}, for a low inclination {\it and} a tidally limited disk
radius), should have an outer disk radius 
$R_{\rm out} \approx 0.46a$ (based on the work of \citet{goo93}).    
However, some systems with a moderate mass ratio 
have exhibited a disk reaching all the way out
to the limit of the Roche lobe radius, e.g. U Gem with a mass ratio 
$q=0.35$ and an expected disk radius of about $\approx 0.42a$  
has been observed in quiescence to exhibit double peaked emission lines   
suggesting an outer disk radius of $0.61a$ \citep{ech07}. 
Consequently, in the present work we consider accretion disk models
with a radius of $\approx 0.3a$ and $\approx 0.6a$.

The outer disk temperature, 
at $R=R_{\rm out} >>R_{\rm in}$,
is given by  
\begin{equation}
T_{\rm out}= 
T_{\rm eff}(x)\left|_{R=R_{\rm out}} = T_0 x^{-3/4}\right|_{R=R_{\rm out}} = 
364,400~{\rm K} 
\left( \frac{ M_{\rm wd}}{M_{\odot}} \right)^{1/4} 
\left( \frac{ \dot{M}}{10^{-6}M_{\odot}{\rm yr}^{-1}} \right)^{1/4} 
\left( \frac { R_{\rm out} } { 10,000~{\rm km}}
\right)^{-3/4},    
\end{equation}
since $x>>1$. The outer disk temperature does not depend on the choice
of the inner disk radius and, as expected, decreases with increasing
$R_{\rm out}$. For $R_{\rm out}=360,000$~km ($\sim 0.6a$)
we obtain $T_{\rm out} \approx 24,800$~K,
and for a disk half this size ($\sim 0.3a$)
we obtain $T_{\rm out}\approx 41,700$~K in the above equation.   

Overall, for a given $M_{\rm wd}$ and $\dot{M}$, the inner disk temperature
depends on the inner disk radius, and the outer disk temperature depends
on the outer disk radius.

\subsection{Dereddening}  

We updated our dereddening software and instead of using the
IUERDAF task UNRED (based on \citet{sav79}, and
which we had been using for {\it IUE} data \citep{god14}),  
we generated our own script using the \citet{fit07} extinction curve
(with $R=3.1$).  

We combine the {\it HST}, {\it GALEX}, and {\it IUE} spectra and remove the  
unreliable spectral portions of {\it GALEX} and {\it IUE} (see Fig.2).  
The interstellar extinction produces  
a strong and broad absorption 
feature centered at 2175~\AA (see Fig.6), due mainly to 
polycyclic aromatic hydrocarbons (PAH) grains \citep{li01}. 
We dereddened the spectrum for different values of 
$E(B-V)$, using the extinction curve from \citet{fit07}. The absorption features
vanishes for $E(B-V)=0.30$, and we therefore adopt this value of the
reddening for T Pyx.  From the visual inspection of the figure alone we infer
an error of 0.05. 
An additional 20 percent error ($\pm 0.06$) 
has to be taken into account when using the 2175~\AA\ bump \citep{fit07}. 
It is worth mentioning, however, that PAHs, the dominant contributor to 
the 2175~\AA\ bump, do not dominate the FUV extinction \citep{li01}, and
three decades ago it was already shown that the 2175~\AA\ bump 
correlates poorly 
with the FUV extinction \citep{gre83}. This is further illustrated 
by the larger sample variance about the mean average Galactic extinction 
curve observed in the shorter wavelength of the FUV \citep{fit07}. 
In the present work, we therefore
adopt $E(B-V)=0.30 \pm 0.08$, and refer the reader to \citet{god14} 
(and to our conclusion section) for an investigation
of how the assumed reddening affect our results, taking into account
that in the literature the reddening toward T Pyx has been 
as low as 0.25 \citep{gil07} and as large as 0.50 \citep{sho13}.  

\begin{figure}
\vspace{-3.cm} 
\figurenum{6}
\plotone{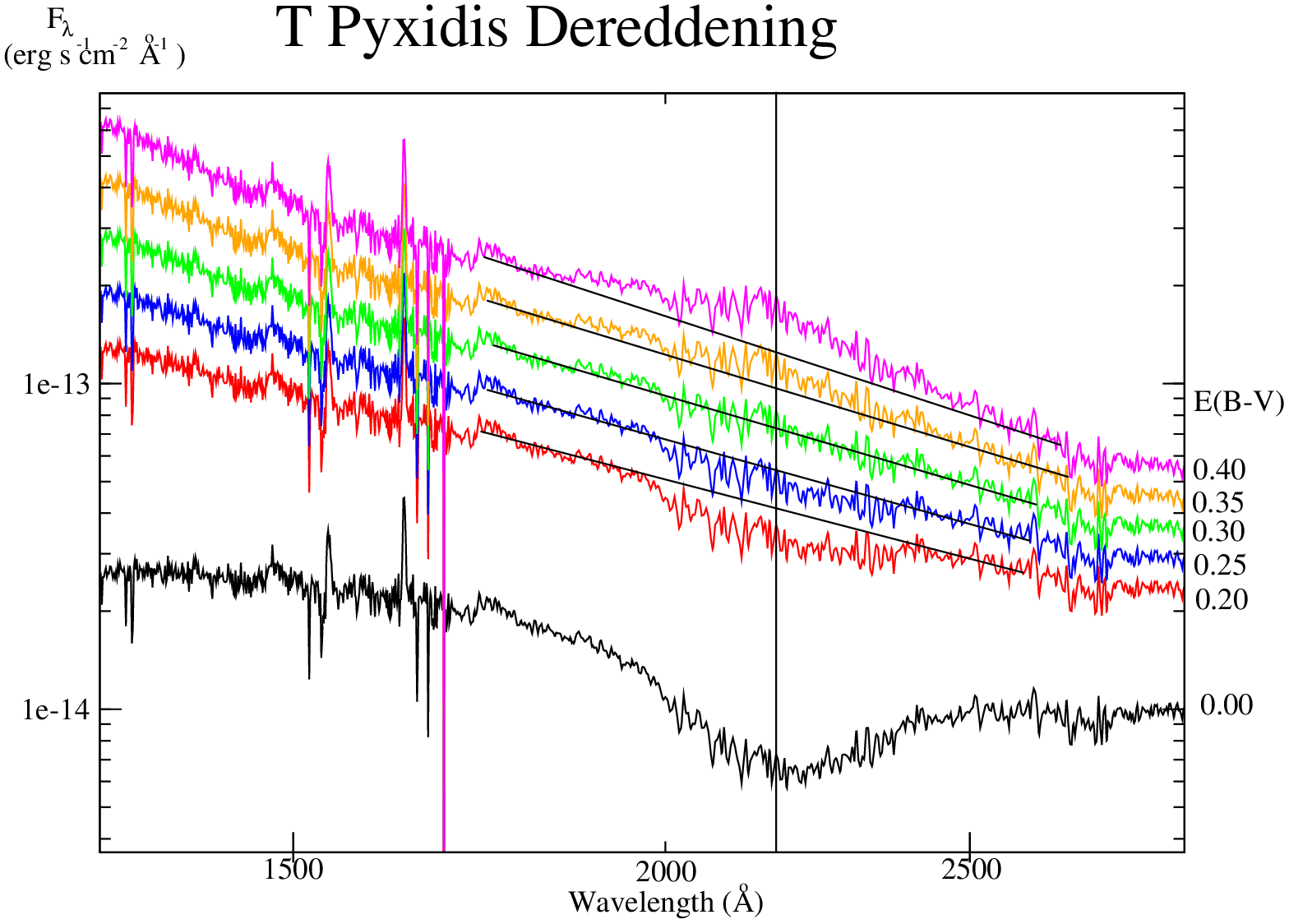}
\caption{
The combined {\it HST} STIS, {\it GALEX}, and {\it IUE} spectrum is presented 
here, where  we have removed the  
unreliable spectral portions of {\it GALEX} and {\it IUE} (see Fig.2).  
The lower spectrum (solid black line) is the 
undereddened spectrum, namely $E(B-V)=0.00$ 
as indicated on the right. 
The interstellar extinction produces 
a strong and broad absorption 
feature centered at 2175~\AA\ (marked with a vertical line). 
We dereddened the (bottom) spectrum for different values of 
$E(B-V)$ 
(again as indicated on the right),
using the extinction curve from \citet{fit07}. The absorption features
vanishes for $E(B-V)=0.30$, and we therefore adopt this value of the
reddening for T Pyx in the present work. In contrast, in 
our previous work \citep{god14} we obtained $E(B-V) = 0.35$
using the \citet{sav79}'s extinction law.  
}
\end{figure}

\clearpage 

\subsection{Optical Wavelengths}

In our current analysis we extend the wavelength coverage into the optical, thereby
now generating disk and WD spectra from 900~\AA\ to 7500~\AA .
Our UV-optical disk modeling has recently been applied in the spectral
analysis of the most extreme RN M31N 2008-12a \citep{dar17}.  

T Pyx was observed by \citet{wil83} in the optical in its pre-outburst state on 
Feb 1st, 1982, at UT 07:07 for 75~min, at the (MDM) McGraw-Hill
Observatory with the 1.3~m telescope and a 2000 channel intensified Reticon 
spectrophotometer. An optical spectrum was obtained in the wavelength
range 3500-7100~\AA , thereby covering the Balmer jump. 
We digitally retrieved the T Pyx optical spectrum from \citet{wil83} 
and scaled it to the pre-outburst (2008) optical spectrum 
obtained by \citet{uth10}, which is itself a flux-calibrated average
of 200 spectra observed with the Magellan telescope 
(4000~\AA\ to 5200~\AA) . 
We use this spectrum to verify and validate our UV disk model and fit. 

\subsection{Parameter Space} 
In the present work we extend the modeling we carried out in \citet{god14} 
over a slightly different region of the parameter space as follows. 
First we assume $E(B-V)=0.30$, while in \citet{god14} we checked how
the results vary for $E(B-V)=$0.25, 0.35, and 0.50. Similarly, 
we currently choose an inclination $i=20^{\circ}$ rather than assuming 
$i=10^{\circ}$ and $i=30^{\circ}$. 
We therefore expect the present results to ``fall'' between our
previous results of $E(B-V)=0.25$ and $0.35$ and 
$i=10^{\circ}$ and $i=30^{\circ}$. 
However, in the present work 
we also check how the results are affected when assuming 
a large inclination, $i=60^{\circ}$,
as suggested by \citet{pat17}. 
As in \citet{god14} we run models for both $M_{\rm wd}=0.70M_{\odot}$
and $M_{\rm wd}=1.35M_{\odot}$. 
Due to size of the disk being limited between $\sim 0.3a$ and $\sim 0.6a$,
we obtain outer disk temperatures of order 
$\sim 40,000$~K   and $\sim 20,000$~K, respectively, while in our
previous work the disk extended to a radius where $T \sim 10,000$~K. 
We choose (as explained earlier) the inner radius of 
the disk to be larger than the WD radius. 

Because the {\it Gaia} DR2 data \citep{pru16,bro18,eye18}
was released after the manuscript
had been submitted and reviewed (and just before it was re-submitted),
we include models for the {\it Gaia} distance of (see section 1.2)
$d = 3.3^{+0.5}_{-0.4}$kpc in the end of the Results and Discussion Sections.

\section{Results} 

As in \citet{god14}, 
in the following we model the co-added post-outburst 
{\it HST} Dec 2012 and Jul 2013 spectrum, 
combined with the pre-outburst co-added {\it IUE} spectra, 
{\it GALEX} spectrum (as shown in Fig.6), together with the 
pre-outburst optical spectrum. The Oct 2015 and Jun 2016 {\it HST} spectra
exhibit a small drop ($\sim 20$\%) in flux in comparison
to the Dec 2012 and Jul 2013 {\it HST} spectra, but are otherwise
almost identical to them. The mass accretion rate corresponding
to the Dec 2012 and Jul 2013 spectra can then simply be derived
by scaling our derived mass accretion rate. 

Since the actual mass accretion rate is not known {\it a priori},
it has to be found by decreasing or increasing $\dot{M}$ until 
a fit at a the distance of 4.8~kpc is found.  
If the decrease or increase in $\dot{M}$ is of the order of 
$\sim 40$\% or smaller 
($\Delta \dot{M}/\dot{M} < 0.4$), 
then it is carried out by linear scaling
the flux of the disk model; if it is larger than that 
($\Delta \dot{M}/\dot{M} > 0.4$),  
then a new disk model is generated.  
Also, some of the parameters that are varied 
do not provide a noticeable change in the results and,  
consequently, we only list the disk models exhibiting
noticeable changes. As the mass accretion rate considered here
is extremely large, the contribution of the WD to the spectrum
is completely negligible and, consequently, cannot be modeled.   
Therefore, we present here only a limited number of models.

We first started with a $1.35 M_{\odot}$ WD with a radius of 2,000~km, 
and a corresponding binary separation $a \approx 600,000$~km.  
Since the mass accretion rate needed to scale to the distance
is very large (of the order of $10^{-6}M_{\odot}$yr$^{-1}$) we
set the inner radius of the disk $R_{\rm in}$ 
to $1.1 R_{\rm wd}$,  $1.2 R_{\rm wd}$,  
and $1.5 R_{\rm wd}$, to mimic a geometrically thick boundary layer.  
These models gave identical results and we decided to adopt 
$R_{\rm in}=1.1 R_{\rm wd}$. We first checked the results for 
an outer disk radius
of the order of $0.3a$ (limited by tidal interaction), 
which, due to the discrete values of the disk's
rings in the model, 
gave $R_{\rm out}=0.27a$. For this model, model \#1 in Table 4,
we obtained a mass accretion rate 
$\dot{M}=1.2 \times 10^{-6}M_{\odot}$yr$^{-1}$ 
with a minimum disk temperature at $R_{\rm out}$ of
$T_{\rm disk}^{\rm out} = 46,700$~K.   
This model is presented in Fig.7 and fits the UV data relatively well
up to a wavelength $\lambda \approx 2,000$~\AA . At longer wavelengths
the model becomes too steep and the spectral slope of the data becomes
flatter with increasing wavelength. Namely, the continuum slope in the
optical is flatter than the NUV continuum slope, which itself is flatter than
the FUV continuum slope.  
The disk model does not
show any sign of the Balmer jump because the disk has 
a temperature $T > 45,000$~K. The observed optical spectrum does not clearly
show the Balmer jump in absorption nor in emission, but does show hydrogen 
emission lines as well as some absorption lines.    

Next, we increased the disk size to a maximum value of $\approx 0.6a$
(about the size of the Roche lobe radius),
which, again due to the discrete values of the disk rings, gave 
$R_{\rm out}=0.59a$. Since this disk surface area is larger, it requires a smaller
mass accretion rate to scale to a distance of 4.8~kpc. For this model,
model \#2 in Table 4, we obtained 
$\dot{M}= 5.6 \times 10^{-7}M_{\odot}$yr$^{-1}$ with a minimum disk
temperature of $T_{\rm disk}^{\rm out}=23,500$~K at $R_{\rm out}$.  
This model, shown in Fig.8, exhibits a smaller discrepancy in the optical
wavelengths and starts to show a ``semblance'' of Balmer jump which is 
not particularly in disagreement with the optical spectrum itself.   
Altogether, the larger disk model (\#2) provides an overall  
better fit than the smaller disk model (\#1). 

We then checked the effect of a higher inclination on the results. We increased the inclination
from $i=20^{\circ}$ to $i=60^{\circ}$ in model \#2 which decreases the overall flux by about
a factor of two. Consequently,
for this new model, model \#3 in Table 4, we had to 
increase the mass accretion rate to $1.36 \times 10^{-6}M_{\odot}$yr$^{-1}$
to obtain the correct fit to the distance. 
This model exhibits wider and shallower absorption lines, 
but as the observed spectrum exhibits mainly emission lines, the fit is
carried out on the continuum.    
This model fit is similar to models \#1 and \#2 in the UV; in the optical, 
however, the model is not as good as model \#2, but it is better than model \#1.  
For all the disk models considered here, we found that 
the increase in the inclination (to $i=60^{\circ}$) 
reduces the flux level by a factor of $\sim 2$ 
(and therefore increasing the mass accretion rate by 
the same factor), due mainly to the
reduced projected emitting area and somewhat to the coefficient of the limb 
darkening. For clarity we only list model \#3 in Table 4.   

\begin{deluxetable*}{ccccccccccc}[b!] 
{\setcounter{magicrownumbers}{-20}}
\tablewidth{0pt}
\tablecaption{T Pyxidis Disk Models Parameters} 
\colnumbers
\tablehead{ 
Model  & $M_{\rm wd}$  & $R_{\rm wd}$ & $a$     & $R_{\rm in}$   & $R_{\rm out}$ & $\dot{M}$   & $T_{\rm disk}^{\rm out}$ & $i$ & d &$\Delta M$ \\  
Number & $(M_{\odot})$ & ($10^3$km)   & $(10^5{\rm km})$ & $(R_{\rm wd})$ & $(a)$   & $(10^{-X}M_{\odot}$yr$^{-1})$ & (10$^3$K)  & (deg) & (kpc) & ($10^{-X}M_{\odot}$) 
} 
\startdata
%
%
\rownumber & 1.35  &  2.0   &  6.0   & 1.1  &  0.27  &   1.20(-6)   &  46.7  &  20   &  4.8 & 5.28(-5)  \\
\rownumber & 1.35  &  2.0   &  6.0   & 1.1  &  0.59  &   5.60(-7)   &  23.5  &  20   &  4.8 & 2.46(-5)  \\
\rownumber & 1.35  &  2.0   &  6.0   & 1.1  &  0.59  &   1.36(-6)   &  26.7  &  60   &  4.8 & 5.98(-5)  \\
\rownumber & 0.70  &  8.5   &  5.0   & 1.1  &  0.30  &   1.92(-6)   &  40.0  &  20   &  4.8 & 8.45(-5)  \\
\rownumber & 0.70  &  8.5   &  5.0   & 1.1  &  0.61  &   1.24(-6)   &  24.2  &  20   &  4.8 & 5.46(-5)  \\
\rownumber & 0.70  &  8.5   &  5.0   & 1.5  &  0.30  &   2.15(-6)   &  40.7  &  20   &  4.8 & 9.46(-5)  \\
\rownumber & 0.70  &  8.5   &  5.0   & 1.5  &  0.59  &   1.36(-6)   &  26.0  &  20   &  4.8 & 5.99(-5)  \\
\rownumber & 1.35  &  2.0   &  6.0   & 7.0  &  0.59  &   7.90(-7)   &  25.7  &  20   &  4.8 & 3.48(-5)  \\ 
\rownumber & 1.35  &  2.0   &  6.0   & 1.0  &  0.30  &   1.03(-7)   &  24.6  &  20   &  2.9 & 4.53(-6)  \\
\rownumber & 1.35  &  2.0   &  6.0   & 1.0  &  0.30  &   1.34(-7)   &  24.6  &  20   &  3.3 & 5.90(-6)  \\
\rownumber & 1.35  &  2.0   &  6.0   & 1.0  &  0.30  &   1.76(-7)   &  24.6  &  20   &  3.8 & 7.74(-6)  \\
\rownumber & 1.35  &  2.0   &  6.0   & 1.0  &  0.60  &   9.00(-8)   &  14.8  &  20   &  2.9 & 3.96(-6)  \\
\rownumber & 1.35  &  2.0   &  6.0   & 1.0  &  0.60  &   1.15(-7)   &  14.8  &  20   &  3.3 & 5.06(-6)  \\
\rownumber & 1.35  &  2.0   &  6.0   & 1.0  &  0.60  &   1.53(-7)   &  14.8  &  20   &  3.8 & 6.73(-6)  \\
\rownumber & 0.70  &  8.5   &  5.0   & 1.1  &  0.30  &   6.80(-7)   &  40.0  &  20   &  2.9 & 2.99(-5)  \\
\rownumber & 0.70  &  8.5   &  5.0   & 1.1  &  0.30  &   8.70(-7)   &  40.0  &  20   &  3.3 & 3.83(-5)  \\
\rownumber & 0.70  &  8.5   &  5.0   & 1.1  &  0.30  &   1.15(-6)   &  40.0  &  20   &  3.8 & 5.06(-5)  \\
\rownumber & 0.70  &  8.5   &  5.0   & 1.1  &  0.60  &   4.21(-7)   &  24.2  &  20   &  2.9 & 1.85(-5)  \\
\rownumber & 0.70  &  8.5   &  5.0   & 1.1  &  0.60  &   5.45(-7)   &  24.2  &  20   &  3.3 & 2.40(-5)  \\
\rownumber & 0.70  &  8.5   &  5.0   & 1.1  &  0.60  &   7.22(-7)   &  24.2  &  20   &  3.8 & 3.18(-5)  \\
\enddata
\tablecomments{In column 4 we list the binary separation $a$, followed by the 
inner and outer radii of the disk $R_{\rm in}$ and $R_{\rm out}$ respectively in 
columns 5 and 6. The outer disk temperature $T_{\rm disk}^{\rm out}$ (reached 
at $R=R_{\rm out}$) is given in column 8 (it is the lowest temperature in the disk).
Models 9 through 20 were computed after completion of the manuscript
to check effect of the just-released {\it Gaia} distance of 
$3.3^{+0.5}_{-0.4}$kpc.
The negative number in parenthesis in columns (7) and (11) denotes
the exponent ``X'' in the unit row.  
The last column indicates the mass of the accreted envelope over 44 years.
} 
\end{deluxetable*}

\begin{figure}
\vspace{-2.cm} 
\figurenum{7} 
\plotone{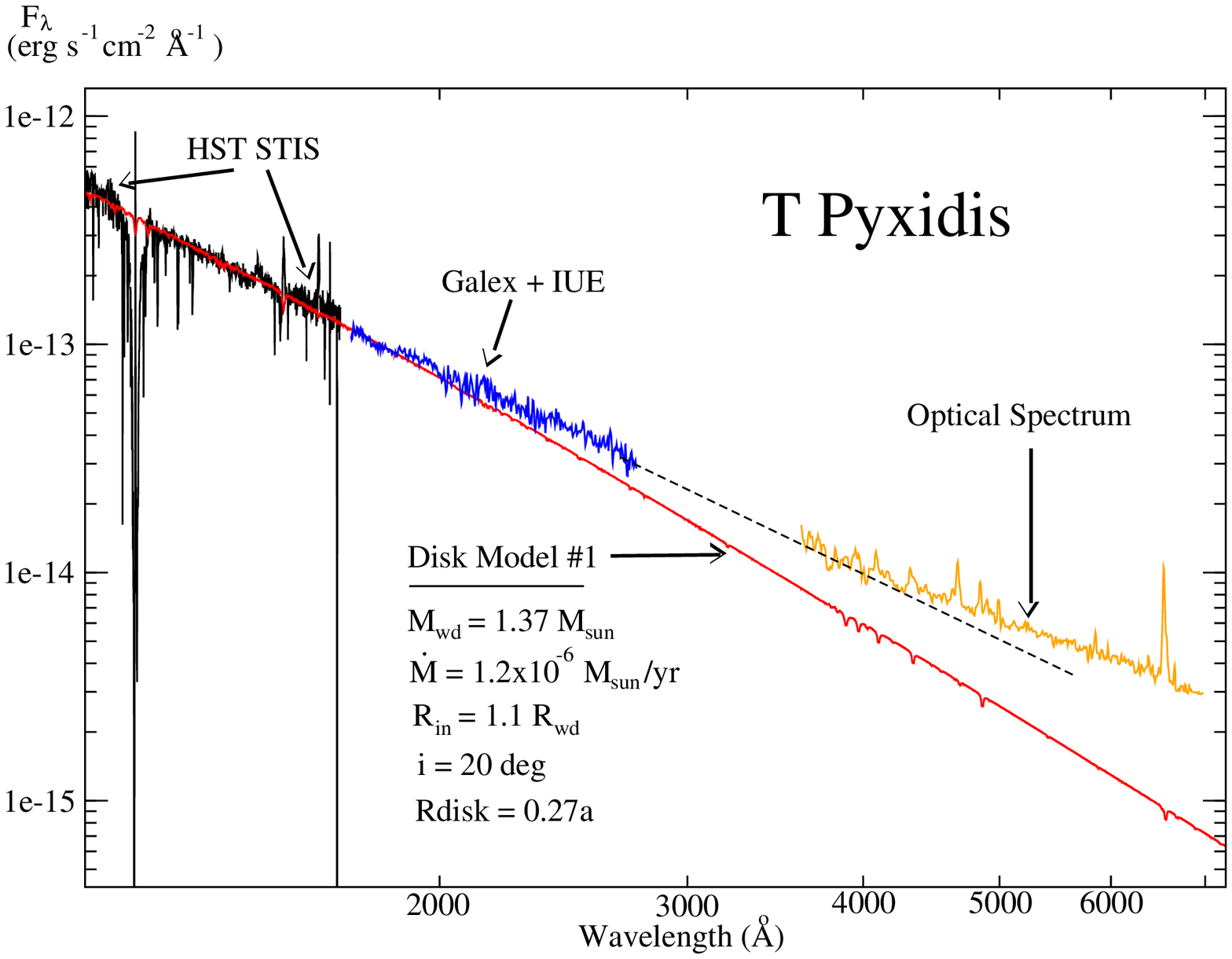}   
\caption{
The {\it HST} STIS, {\it GALEX}, {\it IUE} and optical spectra of T Pyx are presented here 
(color coded) on a log-log scale together with accretion disk
model \# 1 (Table 4; solid red line). The observed 
spectra have been dereddened assuming $E(B-V)=0.30$. 
The optical spectrum is in orange.
The slope of the disk model departs from the data at wavelengths 
$> 2,000$~\AA , namely the slope of the {\it GALEX}+{\it IUE} spectrum (in blue) 
is flatter than the slope of the STIS spectrum (in black), 
and the optical spectrum itself
is even flatter (as shown by the dashed black line for comparison).
The disk model does not
show the Balmer jump because the disk everywhere has 
a temperature $T > 45,000$~K.  
}
\end{figure}

\begin{figure}
\vspace{-2.cm} 
\figurenum{8}
\plotone{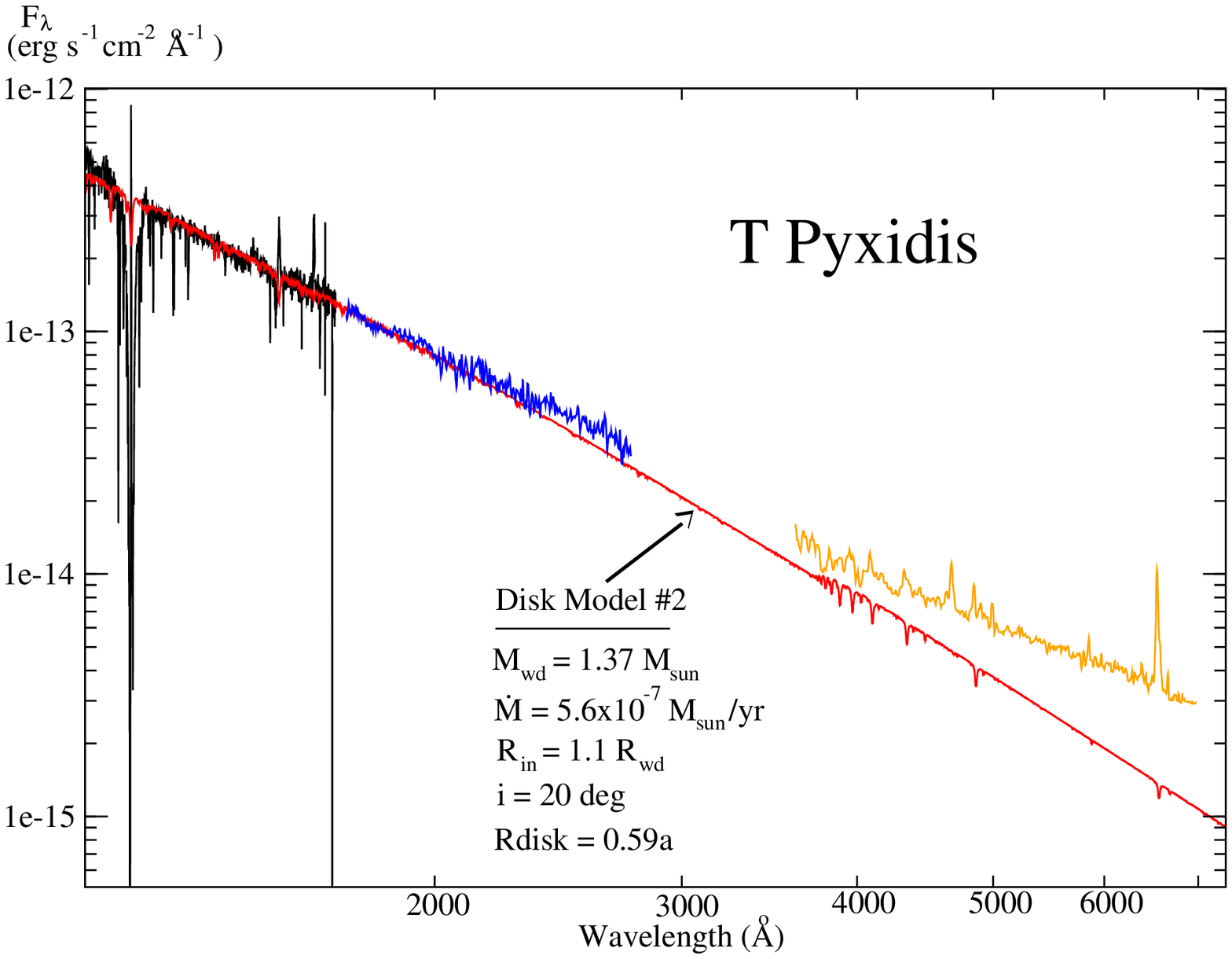}  
\caption{
Same as in Fig.7, but now the disk has been allowed to extend to the
Roche lobe radius, $\approx 0.6a$, where $a$ is the binary separation
(model \# 2 in Table 4). 
Since the disk radius is larger, it has a a larger flux at the same
accretion rate and the resulting $\dot{M}$ needed to fit the data
is lower than in model \# 1 (Fig.7). The outer disk reaches a temperature
as low as 23,500~K and helps flatten the slope of the theoretical spectrum 
resulting in a slightly higher flux in the optical range than model \# 1. Because of its 
lower temperature, the disk model exhibits a Balmer jump, which is
very weak.  
}
\end{figure}

\clearpage

We continued by running models with a lower WD mass, $0.7 M_{\odot}$,
agreeing with the analysis of \citet{uth10}, and adopted a WD radius
of 8,500~km, corresponding to a temperature of $\sim$30,000~K
for such a WD mass \citep{woo95}.  For this primary mass, 
the binary separation shrinks to $\sim$500,000~km. We chose an
inner disk radius $R_{\rm in}=1.1R_{\rm wd}$ and obtained a 
mass accretion rate of $\dot{M}=1.92 \times 10^{-6}M_{\odot}$yr$^{-1}$  
and $\dot{M}=1.24 \times 10^{-6}M_{\odot}$yr$^{-1}$ 
for an outer disk radius of $0.30a$ (model \#4) and $0.61a$ (model \#5) 
respectively. These two models gave the same fits as for the larger WD mass
models \#1 and \#2 and are indistinguishable from them: 
model \#5 fits the optical region better than model \#4 and presents 
an identical fit as model \#2 shown in Fig.8. 
The reason for the similarity of the models lies in the fact that the region
of the disk contributing to the UV has about the same temperature,
as shown in Table 4 by the outer disk temperature reached at $R_{\rm out}$
for models \#4 and \#5. The very inner disk is colder in the $0.7M_{\odot}$
models than in the $1.35 M_{\odot}$ models and choosing a slightly larger
inner disk radius, $R_{\rm in}=1.5 R_{\rm wd}$ increases the mass accretion
rate by about 10\% as shown in Table 4 for models \#6 and \#7. The fits
are, however, identical and one cannot differentiate between the 
$1.1R_{\rm wd}$ and $1.5R_{\rm wd}$ models.    
\\

As stated earlier, the just-released {\it Gaia} parallax gives a distance of 
$3.3^{+0.5}_{0.4}$kpc, significantly smaller than the $4.8-5.0$~kpc
of \citet{sok13,deg14}, thereby demanding further model fits. 
We, therefore, carried out {\it post facto} 12 more model fits:  
for $M_{\rm wd}=1.35M_{\odot}$ and   
$M_{\rm wd}=0.70M_{\odot}$, each for a disk outer radius 
$R_d=0.3a$ and $R_d=0.6a$, and for a distance of $d=$2.9, 3.3, and 3.8~kpc.    
These 12 models are listed in Table 4 (number \#9 through \# 20). 
Overal the {\it Gaia} distance gives a mass accretion of the order
of $10^{-7} M_{\odot}$yr$^{-1}$ for $M_{\rm wd}=1.35 M_{\odot}$  
and $5-7 \times 10^{-7}M_{\odot}$yr$^{-1}$ for 
$M_{\rm wd} = 0.70 M_{\odot}$. 
In the UV range these 12 models provide a fit to the flux continuum slope
as good as models 1 \& 2 in Figs.7 \& 8.    
Our {\it ex post facto} results for the {\it Gaia} distance are 
discussed in the next section. 


\section{\bf{
Discussion
}} 

Our current results agree with our previous analysis \citep{god14}   
that, during its quiescent state, 
T Pyx has a mass accretion rate $\dot{M}$ of the order 
of $10^{-6}M_{\odot}$yr$^{-1}$ for a distance of $4.8$~kpc.  
This is about 10 times larger than the estimate from 
\citet{pat17}, who derived the mass accretion rate from
the period change of the system both in quiescence
and as a result of the eruption. 
The discrepancy between our results and \citet{pat17}'s vanishes
when we considered the {\it Gaia} derived distance of $3.3$~kpc
and a WD mass of $1.35M_{\odot}$. 
However, the discrepancy remains for the low WD mass ($0.7M_{\odot}$)
assumption. 

The exact value of $\dot{M}$ we obtain here depends on the assumed WD mass, 
inclination,  reddening, outer disk radius 
and distance to the system.
We discuss these here below.

\paragraph{\bf{The Optical Range}} 
An important improvement is the inclusion of the optical 
data, which reveal that the optical continuum slope
is flatter than the NUV continuum slope, which itself is flatter than
the FUV continuum slope (see Figs.7 and 8).  
This is contrary to the analysis of \citet{gil07} 
who claimed that the slope becomes steeper at longer wavelengths. 
The discrepancy is likely due to the limited number of data points 
used by \citet{gil07}.  
The slope of the optical continuum is  compatible with an
$\sim$8,000~K stellar atmosphere, but it does not reveal the Balmer
jump (contrary to a stellar atmosphere spectrum).  
Also the emitting area of a 8,000~K component 
would have to be larger than the size of binary
system to produce such a flux at such a low temperature. 
It is, therefore, clear that a significant contribution to the optical
flux comes from the optically thick accretion disk. 

We further present model \#1 and \#2 together with the optical data 
in Fig.9 on a linear scale. 
We display the optical spectrum from \citet{wil83} together with the
optical spectrum from \citet{uth10} to which it was scaled. The similarity
between the emission lines and slope of the continuum between the 
two optical
spectra provides further evidence that the system remains in a state relatively
similar over many years, as already demonstrated by the UV spectra shown in Fig.3.  
The discrepancy between model \#2 and the optical  
data is only $\sim 2 \times 10^{-15}$erg~s$^{-1}$cm$^{-2}$\AA$^{-1}$, 
and the difference between model \#1 and model \#2 is of the same order. 
This is rather small when 
compared to the UV flux which is 10 to 100 times larger, because the
disk emits only a tiny fraction of its energy in the optical band.  
The {\it ex post facto} models based on the {\it Gaia} distance
are very similar to model 1 \& 2, except for models \#12, 13 and 14,
which have a large disk radius ($0.6a$) and a (relatively) lower mass
accretion rate ($\sim 10^{-7}M_{\odot}$yr$^{-1}$). These 3 models 
have an outer disk temperature reaching $\sim 15,000$~K and exhibit
a strong Balmer jump, unlike the optical spectrum. 
Model \#13 is shown in Fig.9 for comparison.
This indicates that the radius of the disk is probably closer to 
$0.3a$ than to $0.6a$. 
The discrepancy in the optical is not  
inconsistent with the data, as this possibly indicates that an additional
component contributing flux in the longer wavelengths is missing from our
modeling.  Similar results were obtained in the spectral modeling of 
the RN M31N 2008-12a \citep{dar17} . 
Since we rule out an 8,000~K component as the source of the optical flux,
the secondary cannot possibly contribute much to the optical, unless
its irradiation by the inner disk increases its temperature significantly.  
We also note that the optical emission from the nebula (extended shell) 
is rather negligible with a continuum flux level of  
$\sim 10^{-17}$erg$^{-1}$s$^{-1}$\AA$^{-1}$ \citep{wil82}, 
two orders of magnitude smaller than the continuum 
flux level of the optical spectra.

\begin{figure}
\vspace{-2.cm} 
\figurenum{9}
\plotone{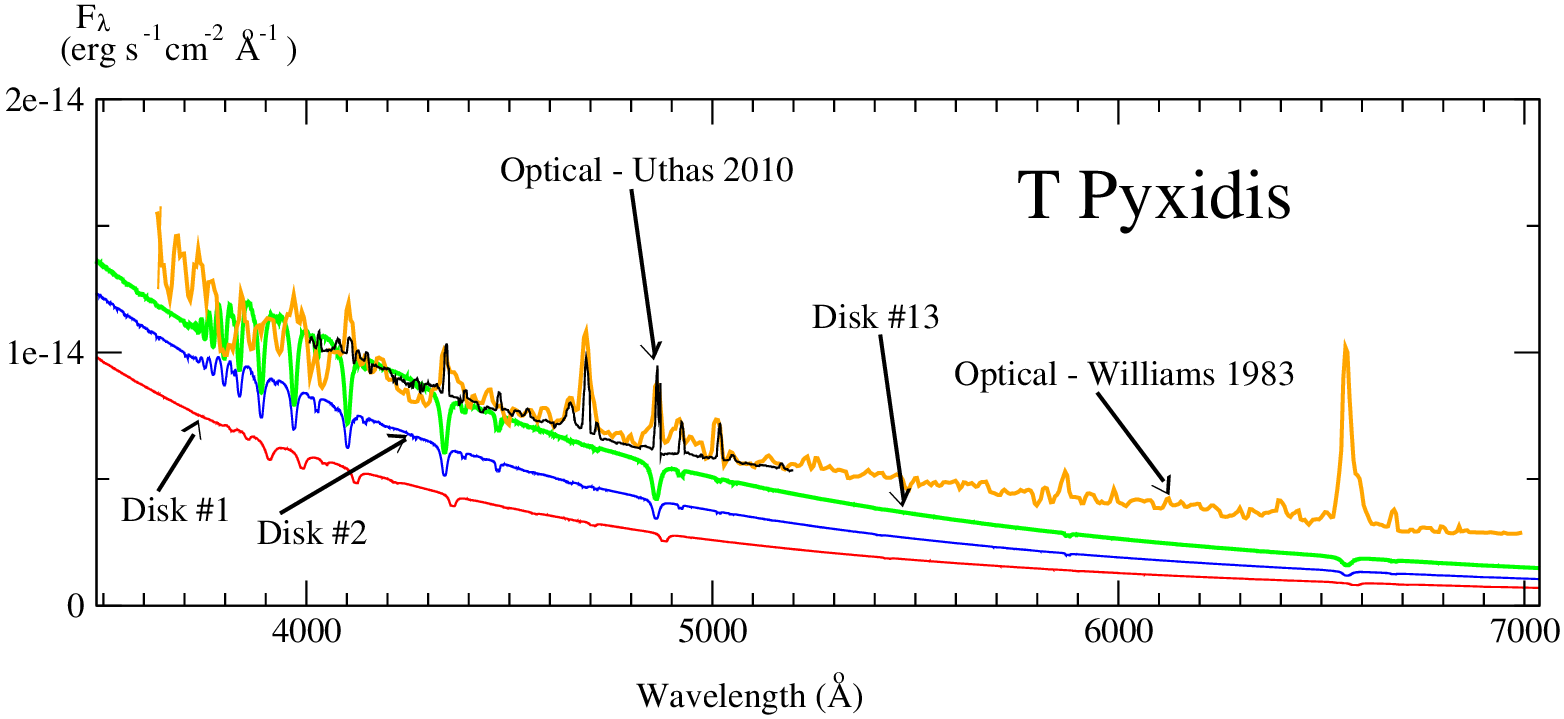} 
\vspace{-4.cm} 
\caption{
Models \#1 (with an outer disk radius of $0.27a$, in blue) 
and \#2 (with an outer disk radius of $0.59a$, in red) 
are shown on a linear scale 
together with the optical spectrum of \citet{wil83} (in orange).
The optical spectrum of \citet{wil83} was scaled to fit the optical
spectrum obtained by \citet{uth10} (in black). The discrepancy between
model \#2 and the optical spectrum could be attributed to an accretion
disk wind \citep{mat15} which would appear here to have a continuum
flux contribution of about 
$2 \times 10^{-15}$erg~s$^{-1}$cm$^{-2}$\AA$^{-1}$.   
The models based the 3.3~kpc Gaia distance give very similar results
except for models \#12, 13 and 14 which have a colder outer disk
displaying a strong Balmer jump. Model \# 13 is shown for comparison.  
}
\end{figure}

\paragraph{\bf{Accretion Disk Wind Contribution}}  
While the best accretion disk models seem to fit the observed spectra
down to about 2000~\AA , the models become too blue in the longer
wavelengths of {\it IUE} and in the optical.  
Often,  CV disk-dominated systems have spectra that cannot be fitted 
with standard disk models, the models are too blue both in the UV 
\citep{pue07} and in the optical \citep{mat15}.  
It was suggested \citep{mat15}  that an accretion
disk wind might be responsible for providing a continuum flux in addition
to the contribution from the optically thick disk, making the
observed overall spectrum redder than that from an optically thick
disk model.  The disk wind is also believed to be responsible for the 
formation of the observed sharp emission lines and for the absence of 
the Balmer jump  when this latter is expected to be in absorption in the 
spectra of disk-dominated CVs. 
It is possible that the discrepancy between our 
optically thick disk model and the observed spectra is due in part to the
lack of an accretion disk wind in our modeling.     
In the present case, however,
the truncation of the outer disk model together with the high 
mass accretion rate generate a theoretical spectrum that does
not exhibit a Balmer jump in agreement with the observed optical
spectrum.


\paragraph{\bf{Long-term behavior}}  
T Pyx seems to have faded by 2 mag since the 1866 nova eruption
\citep{sch05,sch10}, and a look at its AAVSO light curve does indicate
a slight decrease in the $\sim$25 years preceding the 2011 eruption
(Fig.1). The light curve displaying the decline from the 2011 eruption
shows that the system is still possibly declining, as
its magnitude is still increasing. This is further reinforced
by our UV light curve (Fig.5). In addition, it appears that both the optical
(AAVSO) and UV light curves exhibit a drop of the quiescent 
magnitude/flux across  the 2011 eruption: the quiescent magnitude/flux  
reached after the 2011 eruption (say in 2016) is lower than its
pre-outburst value.  
This shows that the mass accretion rate in T Pyx is not only declining 
gradually, but it is also declining after each outburst (or at least
after the last outburst), namely in ``steps''. 
At the present time, this step consists in a 20 percent
drop in the mass accretion rate (Fig.5). 

\paragraph{\bf{X-ray Observations}} 
We have used here optical data to help impose constraints
on the spectral analysis, and, as in \citet{god17b}, we now wish to use 
X-ray data to further constrain and interpret the results of the spectral analysis. 
The main characteristics of the X-ray observations of T Pyx obtained
during quiescence \citep{bal10} is the fact that the X-ray 
luminosity is orders of magnitude smaller than the expected disk
luminosity and that it originates in the shocked nebular material
rather than in the inner accretion disk or boundary layer.  

In the case of a 
massive WD ($1.35M_{\odot}$) accreting at a rate of the order 
of $10^{-6}M_{\odot}$yr$^{-1}$ (models \#1, 2, \& 3), 
the temperature in the inner disk
(even without a boundary layer) reaches a maximum of 
$\sim$500,000~K, 
and drops to $\sim$350,000~K for 
$\dot{M} \approx 10^{-7} M_{\odot}$yr$^{-1}$. 
Such a high temperature component is expected to show 
in the soft X-ray band,and 
the boundary layer temperature would possibly be of the same order 
of magnitude. 
However, X-ray observations 
of T Pyx \citep{gre02,sel08,bal10} do not provide supporting
observational evidence  for such an X-ray source scenario.  

In contrast, if we consider the $0.7 M_{\odot}$ WD mass
disk models, the resulting maximum temperature in  
the inner disk reaches $\sim$170,000~K at $R=1.36 R_{\rm in}$, 
dropping to $\sim$150,000~K at $R=2.10 R_{\rm in}$, and $\sim$100,000~K 
at $R=4.00 R_{\rm in}$. This is for $R_{\rm in}=1.1R_{\rm wd}$
(models \# 4 \& 5), and the temperature drops an additional  20 
percent for $R_{\rm in}=1.5 R_{\rm wd}$ (models \# 6 \& 7).  
A slightly lower temperature is reached for models \#15 thru \#20. 
As such the disk should emit in the EUV rather than soft X-ray,
with the inner disk peaking around 300~\AA .  
The optically thick boundary layer itself could be geometrically thick  
(even if it is dynamically thin, \citep{god95}),
with a similar temperature ($\sim 10^5$~K). Such a scenario would explain the
low X-ray luminosity as all the energy would be radiated
in the EUV. Many CVs accreting at a high mass accretion rate have similar
X-ray characteristics, namely no sign of an optically thick boundary 
layer \citep{fer82}, but instead a very faint hard X-ray emission 
with a luminosity orders of magnitude smaller than the disk luminosity 
\citep{mau95,van96,bas05}.  
The consequence of having a strong EUV source is that the radiation is expected to interact
strongly with the ISM and hence the EUV luminosity 
is greatly reduced. EUV sources are mostly detectable out to a distance
of only a few hundred parsecs and at the shortest wavelengths 
\citep{bar14}\footnote{
The First and Second {\it Extreme Ultraviolet
Explorer (EUVE)} Source Catalogs \citep{bow94,bow96} reveal a dozen 
magnetic CVs and few disk systems (IX Vel, SS CYg, VW Hyi)  
with a very low Galactic extinction in their direction ($E(B-V) < 0.1$) 
or very nearby ($d \sim 200$~pc at most).  
The only exception is nova V1974 Cyg at $\sim 2.5$~kpc  which 
underwent a classical nova explosion in early 1992 \citep{col92a,col92b} 
and was observed less than a year later with EUVE. 
}. 

Therefore, one option is that the mass of the WD is  
$0.7 M_{\odot}$ \citep{uth10}, 
and that the accretion disk emits mostly in the EUV, and that
the EUV radiation cannot reach us. The problem with this scenario is that 
the X-ray emission during outburst argues against a low mass WD, 
and a $0.7M_{\odot}$ WD accreting at a high rate grows to large radii with
no outburst \citep{new14}.

We note, however, that in order for a $1.35M_{\odot}$ WD accreting 
at a high rate ($\sim 10^{-7}-10^{-6}M_{\odot}$yr$^{-1}$) 
to not exhibit X-ray emission, the inner radius of the disk has to 
be much larger than the anticipated WD radius (see equation 3).   
In Table 4 we list such
a model, \# 8.
In that model, we increased the inner disk radius till the maximum
temperature in the inner disk drops to 150,000~K. For this to happen
we find a radius of $\sim$14,000~km, namely 7 times larger than the expected WD radius
(2,000~km for a $1.35M_{\odot}$). For an outer disk radius of 0.59a, the 
disk gives a mass accretion rate of $7.9 \times 10^{-7}M_{\odot}$yr$^{-1}$. 
The fit is as good as model \# 2.
Similarly, if we consider model \# 13 (similar to model \#8 but
adjusted to the {\it Gaia} distance), we find that in order to
reduce the inner disk temperature from 350,000~K to 150,000~K,
we need to increase the inner disk radius to 8,000~km, namely
4 times larger than the expected WD radius.
However, we find no physical explanation  
for the inner disk radius to be so large (T Pyx is not an intermediate polar
and we do not consider disks departing from the standard model).   

\paragraph{\bf{Accreted Mass versus Ejected Mass}}  
We recapitulate our results in Figs.10a, b, \& c, where we  
compare the accreted envelope mass obtained from our analysis
with the estimates of the ejected envelope mass from the literature.  
Our previous accreted mass estimate for $E(B-V)=0.25$, 0.35 and 0.50
(all in black) were computed in \citet{god14} for $i=10^{\circ}$ and $30^{\circ}$. 
Our new results investigating the effects of the size of the disk 
for $E(B-V)=0.30$ and $i=20^{\circ}$ are in red, and as expected 
fall between our previous $E(B-V)=0.25$ and $E(B-V)=0.35$ results.   
The ensuing  accreted envelope is more massive when assuming a 
smaller disk radius. 
Our new results investigating the effects of a large
inclination are in blue and indicate that the accreted envelope 
mass increases by a factor of two when increasing the assumed
inclination from $20^{\circ}$ to $60^{\circ}$.  
These results confirm that 
if we assume a distance of 4.8~kpc 
for most of the parameter space  
the accreted envelope is larger than the ejected one and the
WD should increase its mass with time.

Howevere, if the WD mass in T Pyx
is massive and if the distance to the system is $\sim 3.3$~kpc,
then T Pyx is losing more mass than it is accreting.

\clearpage

\begin{figure*}
\vspace{-7.0cm} 
\figurenum{10a}
\plotone{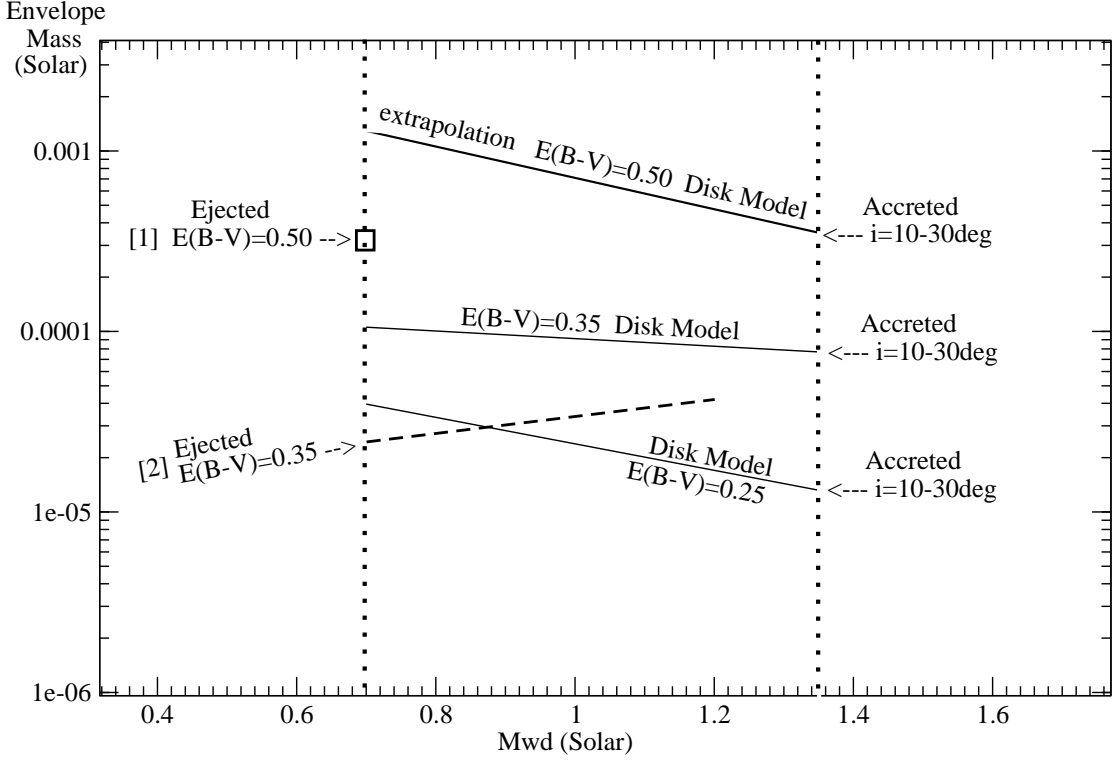}
\caption{
The ejected and accreted envelope masses (Y-axis, in units of 
Log($M_{\odot}$); per nova eruption) 
as a function of the assumed WD mass (X-axis, 
in units $M_{\odot}$), 
based on the 2011 eruption and modeling of the quiescent spectra.
The assumed $0.7M_{\odot}$ and $1.35M_{\odot}$ WD masses are indicated
with the vertical dotted lines. 
The three solid black lines are the accreted envelope mass estimates from 
the UV spectral modeling we carried out in \citet{god14}
assuming different values for the reddening 
and for a system inclination of $10^{\circ}$ and $30^{\circ}$   
(as marked on the figure). 
The first ejected envelope mass [1] (black square symbol in the upper left) 
is the mass loss maximum estimate of \citet{nel14} assuming a reddening of 
$E(B-V)=0.50$ \citep{sho13} and $M_{\rm wd}=0.7M_{\odot}$. 
This ejected mass is to be compared
to our accreted envelope mass for the same reddening (the upper
solid black line) and same WD mass. 
The second ejected envelope mass [2] 
(dashed black line) is from \citet{pat17} and was computed assuming 
$E(B-V)=0.35$ for a range of WD masses. This ejected mass estimate is
to be compared to our accreted envelope mass with the same reddening,
namely the (almost horizontal) solid black line in the middle of the graph. 
Our estimate of the accreted envelope mass for a low reddening 
$E(B-V)=0.25$ (as derived by \citet{gil07}) is the lower solid black line.  
Our previous results (assuming $d=4.8$~kpc) 
indicate that the accreted mass is larger than the
ejected mass and the WD mass is possibly growing with time. 
}
\end{figure*}

\clearpage

\begin{figure*}
\vspace{-7.0cm} 
\figurenum{10b}
\plotone{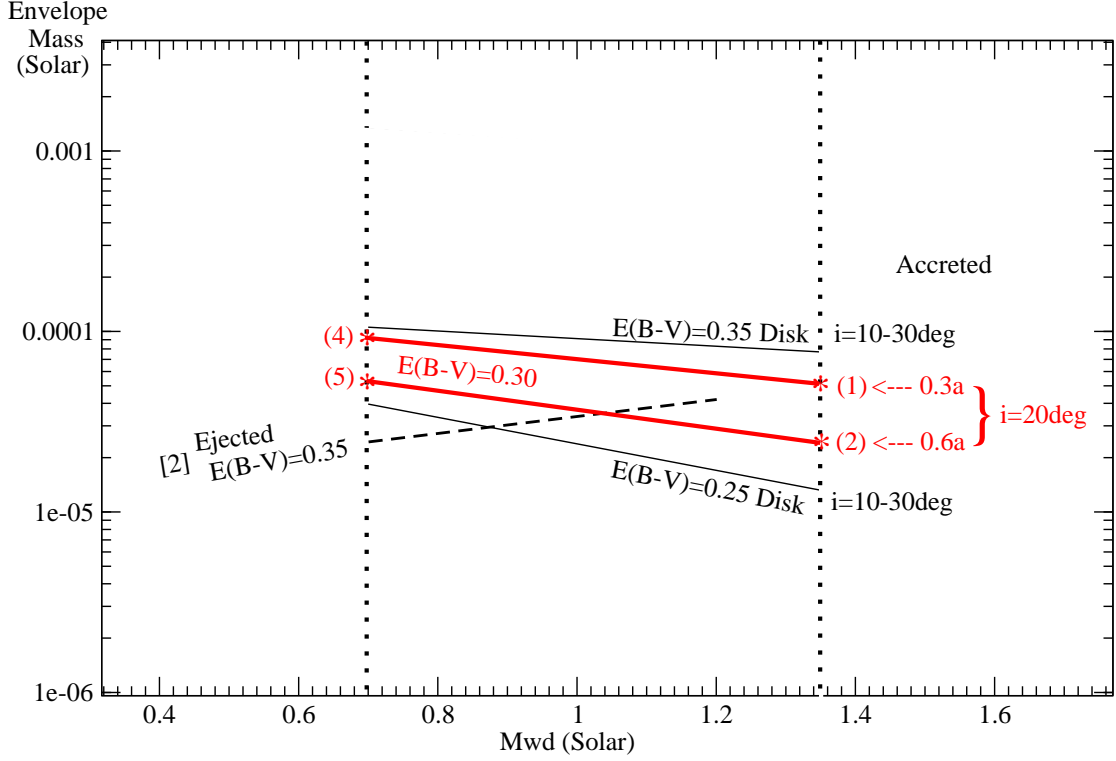}
\caption{
The ejected and accreted envelope masses 
as a function of the WD mass as in Fig.10a.  
We have now added our new results assuming a distance of 4.8~kpc 
(two thick red lines),
using a system inclination of $20^{\circ}$ and 
dereddening the spectra for $E(B-V)=0.30$.  
The models from Table 4 are indicated with
an asterisk * and their respective number (1), (2), (4), and (5)
and the thick solid red lines show the interpolation 
between the $0.7 M_{\odot}$ and $1.35 M_{\odot}$ models.  
For a disk tidally truncated at $0.3a$ (upper thick red line) 
the accreted mass envelope will be larger than for disk extending
all the way to the Roche lobe ($0.6a$, lower thick red line). 
Consistent with our previous results, the present $E(B-V)=0.30$ results  
(the two thick red lines) fall between our $E(B-V)=0.25$ and 0.35 models 
(the two thin black lines).  
For clarity we have removed the $E(B-V)=0.50$ ejected and accreted
mass envelopes from this graph. 
}
\end{figure*}

\clearpage

\begin{figure*}
\vspace{-7.0cm} 
\figurenum{10c}
\plotone{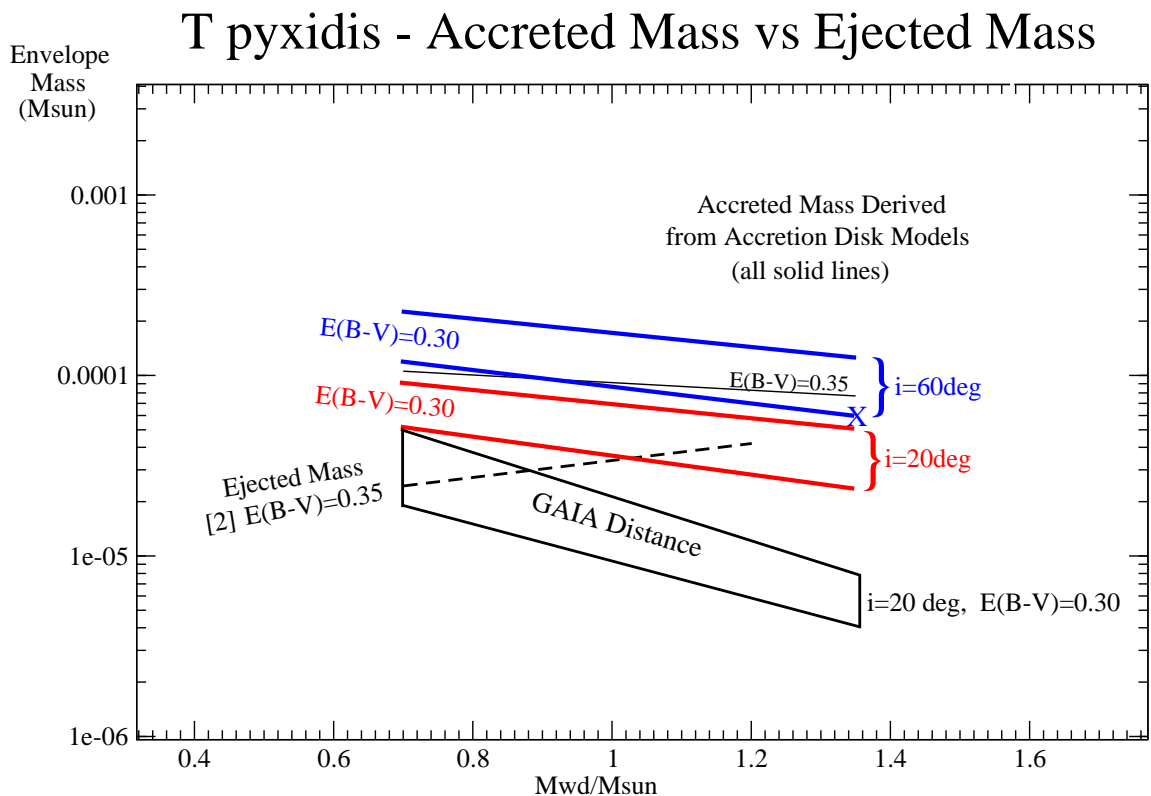}
\caption{
The ejected and accreted envelope masses 
as a function of the WD mass as in Fig.10b.  
We have now added two thick solid blue lines   
for a system inclination of $60^{\circ}$ (as marked on the right).  
Model \# 3 from Table 4 is marked with a large ``X'' sign.
For clarity we have not marked models \# 6 and 7 
and we have removed the vertical dotted lines delimiting 
the WD masses $0.7M_{\odot}$ and $1.35 M_{\odot}$. 
Increasing the inclination from 20$^{\circ}$ (in red) 
to 60$^{\circ}$ (in blue) results in an accreted envelope mass  
twice as large. 
The results presented in Figs. 10a, b, and c, were all obtained assuming
a distance of 4.8~kpc.  
Our {\it ex post facto} results taking into account the shorter
{\it Gaia} distance of $3.3^{+0.5}_{-0.4}$~kpc have been marked
with the thick black line parallelogram in the lower part of the 
graph as the closer distance reduces the mass accretion rate. 
}
\end{figure*}

\clearpage

\section{\bf{
Summary and Conclusion
}} 

We have carried out a UV and optical spectral analysis of recent 
{\it HST} observations of T Pyx in its quiescent and late declining states 
with an improved version of our accretion disk models.  Using archival 
{\it IUE} data, optical data and results from X-ray observations, we 
show that the results heavily depends on the assumed parameters. 
We summarize our findings as follows.

\begin{enumerate} 

\item 
Both the optical data (as of late 2017) and the UV data (as of mid-2016) 
indicate that T Pyx is still declining from its 2011 outburst. 

\item 
Both the UV and optical data show that during the pre-outburst 
quiescent phase the system was in a relatively steady state with the
same continuum shape exhibiting a fluctuation of only 9\% in the continuum
flux level, which can possibly be attributed to orbital modulation.  

\item 
The UV clearly reveals a drop in the accretion rate {\it across } the
outburst, namely when comparing
the pre-outburst spectra to the late decline post-outburst spectra.     
This drop is confirmed by the AAVSO light curve (Fig.1), which also exhibits 
a drop across  the outburst.   
We suggest that the system possibly experiences such a drop, 
after each recurring nova eruption, which might contribute to the
2 mag fading since 1866.  

\item
The previously accepted distance of $4.8$~kpc gives a large mass
accretion rate ($\dot{M} \approx 10^{-6}M_{\odot}$yr$^{-1}$, ten times
larger than the estimate of \citet{pat17}) 
yielding to an accumulated envelope mass of $\approx 2.5-9.5 \times 10^{-5}
M_{\odot}$ (models 1-8). 
The mass ejected during the eruption is estimated to be 
$\sim 3 \times 10^{-5}M_{\odot}$ \citep{pat17}. 
This would  imply that the T Pyx WD mass
increased from 1966 (post-outburst) to 2011 (post-outburst), in agreement  
with \citet{new14} who predict that, irrespective of the actual mass of
the WD, the WD will continue to grow in mass. 
However, 
the latest distance estimate from {\it Gaia} ($\sim 3.3$~kpc), 
gives an accumulated envelope mass
as low as $\approx 4 - 8 \times 10^{-6}M_{\odot}$ (models 9-14). 
Only if the WD in T Pyx is as low as $0.7M_{\odot}$ 
does the accumulated envelope mass reaches $\approx 2-5
\times 10^{-5} M_{\odot}$ (models 15-20).

\item
Our results indicate that {\it if} the T Pyx's inclination is large, 
then the mass accretion
rate must be larger by at least a factor of two. 
A larger inclination might, as a result of an inflated disk 
self-obscuration,  
explain why no soft X-ray emission is observed from the inner disk   
and how the UV flux varies by $\pm $9\%. In this case the system would emit
soft X-rays but it wouldn't be observed.  

\item
If, however, the system is not emitting any soft X-rays at all, it is possible
that the WD mass might be as small as $0.7 M_{\odot}$ 
(\citet{uth10}, in disagreement with the theory \citet{sta85,new14}), 
with an inflated radius, 
and the disk inner radius might be $R_{\rm in} \approx 1.5R_{\rm wd}$
to explain the low inner disk/boundary layer temperature emitting
in the EUV rather than in the soft X-ray band.  Such EUV radiation
would be absorbed by the ISM and would not be observable 
at a distance of 4.8~kpc. For a $1.35 M_{\odot}$ WD accreting at a high 
rate to peak in the
EUV (rather than in the X-ray) band, the inner radius of the disk would
have to be 4-7 times larger than the actual radius of the $1.35M_{\odot}$ WD. 


\item 
The reddening toward T Pyx remains largely unknown. 
A lower value of $E(B-V) \approx 0.25-0.35$ is obtained
when using the 2175~\AA\ PAH bump, which correlates poorly with the FUV
extinction, \citep{gre83}. 
A value twice as large, $E(B-V)\approx 0.50$,  
is obtained using the diffuse interstellar bands, however, $E(B-V)$ 
versus the $W_{\lambda}$(5780.5) has a rather large scatter when
considering the data points individually \citep{fri11}. Both techniques
have their limitations and the results have to be considered for all values 
$E(B-V) = 0.35 ^{+0.15}_{-0.10}$ (Fig.10). 

\end{enumerate} 
 
To conclude, we point out two possible scenarios emerging 
from our analysis as follows. 

(i) The WD is massive, $1.35M_{\odot}$, with a small radius, 
and due to the high accretion rate the inner disk emits soft X-rays which 
are blocked by the thick portion of the disk possibly 
viewed at a large inclination
$i=60^{\circ}$. There is no restriction on the reddening.  

(ii) The inner radius of the disk is large ($\sim 10-15,000$~km,
due to either a  small WD mass $0.70M_{\odot}$, or the truncation of 
the inner disk), the inclination is 
$20 \pm 10^{\circ}$, and the reddening is probably $E(B-V) \approx 0.25-0.30$
(to keep $\dot{M}$ to a value of $\sim 10^{-6}M_{\odot}$yr$^{-1}$). 
The inner disk and boundary layer have a
maximum temperature $\sim 150,000$~K peaking in the EUV. 
At a distance of a few kpc the EUV radiation is absorbed by the ISM.  

In both cases the disk is large, and the UV light is modulated by the 
orbital motion as matter from the L1-stream overflow the disk rim and 
further masks parts of the inner disk. 
Most importantly, the newly derived distance from the {\it Gaia} DR2
data, for the most plausible values of the WD mass ($1.35M_{\odot}$)
and reddening ($\approx 0.3$), imply a mass accretion rate of the order
of $10^{-7}M_{\odot}$yr$^{-1}$, such that the accumulated envelope mass
is actually smaller than the mass ejected during the eruption, indicating
that the WD mass does not grow in mass. However, the Gaia distances for
variable/binary stars have to be taken with some reservations 
\citep{eye18}.

\acknowledgements 

We wish to thank the referee for her/his prompt report which helped improve
the figures and correct an omission in the {\it IUE} dataset.  
PG is pleased to thank William (Bill) P. Blair at the Henry Augustus Rowland
Department of Physics \& Astronomy at the Johns Hopkins University, Baltimore,
Maryland, USA, for his kind hospitality. 
It is a pleasure to thank Glen Williams, from Central Michigan University,
Mount Pleasant, Michigan, USA, and Christian Knigge \& Danny Steeghs, 
from the University of
Southampton, Southampton, UK, for giving us permission  
to (digitally) extract and use their optical spectra.  
This research was supported by HST guest observer grants 
GO-12799, GO-12890, and GO-14111, (all with PI E.M. Sion)  
to Villanova University.  
The analysis of the archival {\it IUE} and {\it GALEX} data was supported by funding   
from the National Aeronautics and Space Administration (NASA) under grant
number NNX17AF36G (PI Godon) issued through the Office of Astrophysics and DATA Analysis
Program (ADAP) to Villanova University. 
SS acknowledges partial support from {\it HST} and NASA grants to ASU. 
In our research, we made use of online data
from the AAVSO {\it International Database} 
and we are thankful to the AAVSO and its members 
worldwide for their constant monitoring of CVs
and for making their data public.  
Special thanks 
to the British Astronomical Association Variable Star Section
(BAAVSS) for their contribution to T Pyx's AAVSO data.  
This work has made use of data from the European Space Agency (ESA)
mission {\it Gaia} {\url{https://www.cosmos.esa.in/gaia}}, processed
by the {\it Gaia} Data Processing and Analysis Consortium (DPAC, 
{\url{https://www.cosmos.esa.in/web/gaia/dpac/consortium}}). 
Funding for the DPAC has been provided by national institutions, in particular
the institutions participating in the {\it Gaia} Multilateral Agreement.

\facilities{{\it HST}(STIS, COS), {\it GALEX}, {\it IUE}, 
MDM:McGraw-Hill 1.27m Telescope, AAVSO, GAIA}

\software{CalSTIS pipeline (v3.4), 
CalCOS pipeline (v3.1.8), IRAF (v2.16.1, \citet{tod93}), 
Tlusty (v203) Synspec (v48) Rotin(v4) Disksyn (v7) \citep{hub17a,hub17b,hub17c}, 
FORTRAN (77), PGPLOT (v5.2), Cygwin-X (Cygwin v1.7.16),
xmgrace (Grace v2), XV (v3.10), WebPlotDigitizer (v3.9).} 
\\ \\

Patrick Godon \url{https://orcid.org/0000-0002-4806-5319} 

Sumner Starrfield \url{https://orcid.org/0000-0002-1359-6312}

\end{document}